%% file: main.tex
\documentclass{vldb_nocopyright}
\usepackage{balance}  
\usepackage{url}
\usepackage{fullpage}
\usepackage{graphicx}
\usepackage{xspace}
\usepackage{amsmath}
\usepackage{amssymb}
\usepackage{color}
\usepackage{float}
\usepackage{listings}
\usepackage{enumerate}
\usepackage{multirow}
\usepackage{hyperref}
\usepackage[export]{adjustbox}
\usepackage{graphics}
\usepackage{times}
\usepackage{inconsolata}
\usepackage{tabularx}
\usepackage[noend]{algpseudocode}
\usepackage{algorithm}
\usepackage{verbatim}
\usepackage{mathtools}
\usepackage{caption}
\usepackage{subcaption}
\usepackage{xspace}
\usepackage{mathrsfs}
\usepackage{placeins}
\usepackage{tikz}
\usepackage{listings}
\usepackage{booktabs}
\usepackage{enumitem}
\usepackage{fixltx2e}

\newenvironment{packed_item}{
\begin{itemize}[leftmargin=0.5em,itemindent=15pt,topsep=2pt]
   \setlength{\itemsep}{1pt}
   \setlength{\parskip}{0pt}
   \setlength{\parsep}{0pt}
}
{\end{itemize}}

\DeclareRobustCommand*\circled[1]{\tikz[baseline=(char.base)]{
    \node[shape=circle,draw,inner sep=1pt] (char) {#1};}}

\newcommand{\comm}[2]{{#1}: [{#2}]}
\renewcommand{\comm}[2]{} 

\newcommand{\magda}[1]{{\color{blue}\comm{magda}{#1}}}
\newcommand{\parmita}[1]{{\color{green}\comm{parmita}{#1}}}

\newcommand{\alvin}[1]{{\color{magenta}\comm{alvin}{#1}}}
\newcommand{\ariel}[1]{{\color{teal}\comm{ariel}{#1}}}

\newcommand{\figlabel}[1]{\label{#1}}
\newcommand{\seclabel}[1]{\label{#1}}

\newcommand{\lstlabel}[1]{\label{#1}}

\newcommand{\secref}[1]{\autoref{#1}}  
\newcommand{\figref}[1]{\autoref{#1}}     
\newcommand{\lstref}[1]{\autoref{#1}}

\newcommand{\eg}{{\em e.g.}}
\newcommand{\ie}{{\em i.e.}}

\newcommand{\ea}{{\em et al.}}
\newcommand{\aka}{{\em a.k.a.}\xspace}
\newcommand{\neurostep}[1]{Step~\circled{#1}\textsubscript{N}}
\newcommand{\astrostep}[1]{Step~\circled{#1}\textsubscript{A}}

\newcolumntype{L}{>{\arraybackslash}m{2cm}}
\definecolor{dkgreen}{rgb}{0,0.6,0}
\definecolor{gray}{rgb}{0.5,0.5,0.5}
\definecolor{mauve}{rgb}{0.58,0,0.82}

\begin{document}

\title{Comparative Evaluation of Big-Data Systems on Scientific Image
  Analytics Workloads}
\subtitle{Experiments and Analysis}

\numberofauthors{1}

\author{
\alignauthor \large
Parmita Mehta, Sven Dorkenwald, Dongfang Zhao, Tomer Kaftan, Alvin Cheung\\
Magdalena Balazinska, Ariel Rokem, Andrew Connolly, Jacob Vanderplas,
Yusra AlSayyad \\
\affaddr{University of Washington} \\
}
\maketitle

\begin{sloppypar}

\input{abstract}
\input{intro}

\input{systems}
\input{benchmark}
\input{setup}
\input{experiment}
\input{futurework}
\input{relatedwork}
\input{conclusion}

\end{sloppypar}

\scriptsize
\bibliographystyle{abbrv}
\bibliography{header,ref}

\end{document}

%% file: abstract.tex
\begin{abstract}

  Scientific discoveries are increasingly driven by analyzing
  large volumes of image data. Many new libraries and specialized
  database management systems (DBMSs) have emerged to support such
  tasks. It is unclear, however, how well these systems support
  real-world image analysis use cases, and how performant are the
  image analytics tasks implemented on top of such systems. In this
  paper, we present the first comprehensive evaluation of large-scale
  image analysis systems using two real-world scientific image data
  processing use cases. We evaluate five representative 
  systems (SciDB, Myria, Spark, Dask, and TensorFlow) and 
  find that each of them has shortcomings that
  complicate implementation or hurt performance. Such shortcomings
  lead to new research opportunities in making large-scale
  image analysis both efficient and easy to use.

\end{abstract}

%% file: intro.tex
\section{Introduction}
\seclabel{intro}



With advances in data collection and storage technologies, 
data analysis has become widely accepted as the fourth paradigm of science
\cite{fourthparadigm}.
In many scientific fields, an increasing portion of this data is 
images~\cite{imagedata1, imagedata2}.
It is thus crucial for big data systems\footnote{\small{In this paper,
    we use the term ``big data system'' to describe any DBMS or
    cluster computing library that provides parallel processing
    capabilities on large amounts of data.}} to provide a scalable and
efficient means to store and analyze such data, and support programming
models that can be easily utilized by domain scientists (\eg,
astronomers, physicists, biologists, {\em etc}).

As an example, the Large Synoptic Survey Telescope (LSST) is a
large-scale international initiative to build a new telescope for
surveying the visible sky~\cite{lsst} with plans to collect 60
petabytes of images over 10 years. In previous
astronomy surveys (\eg, the Sloan Digital Sky Survey
(SDSS)),
the collected images were processed by an expert
team of engineers on dedicated servers, with results distilled into
textual catalogs for other astronomers to analyze. In contrast, one of
the goals of LSST is to broaden access to the
collected images to astronomers around the globe, and enable them to
run analyses on the images themselves. 
Similarly, in neuroscience, several large collections of
imaging data are being compiled. For example, the UK biobank will
release Magnetic Resonance Imaging (MRI) data from close to 500k human
brains (more than 200 TB) for neuroscientists to
analyze~\cite{miller2016multimodal}. Multiple other initiatives are
similarly making large collections of image data available to
researchers~\cite{abcdstudy,jernigan2016pediatric,2005astro.ph.10346T}.


Such use cases emphasize the need for effective tools to support the
management and analysis of image data: tools that are efficient, 
scale well, and are easy to program without requiring
deep systems expertise to deploy and tune.



Surprisingly, there has been only limited work from the data
management research community in building tools to support large-scale
image \textit{management and analytics}.  Rasdaman~\cite{rasdaman} and
SciDB~\cite{scidb} are two well-known DBMSs that specialize in the storage
and processing of multidimensional array data and they are a natural
choice for implementing image analytics. Most other work
developed for storing image data targets predominantly image storage
and retrieval based on keyword or similarity searches 
\cite{faloutsos:94,carson:99,chaudhuri:04}. Recently developed parallel 
data processing
libraries such as Spark~\cite{spark} and TensorFlow~\cite{tensorflow:2015}
can also support image data analysis. Hence, the key questions that
we ask in this paper are: How well do these existing 
big data systems support the management and
analysis requirements of real scientific workloads? Is it 
easy to implement large-scale analytics using these systems?
How efficient are the resulting applications that
are built on top of such systems? Do they require deep technical
expertise to optimize? 



In this paper, we present the first comprehensive study of the issues
mentioned above. Specifically, we choose five big data systems that
encompass all major paradigms of parallel data processing: a
domain-specific DBMS for multidimensional array data (SciDB~\cite{scidb}), a
general-purpose cluster computing library with persistence
capabilities (Spark~\cite{spark}), a traditional parallel
general-purpose DBMS (Myria~\cite{halperin:14,wang:17}), and 
a general-purpose (Dask~\cite{dask:2015}) and domain-specific
(TensorFlow~\cite{tensorflow:2015}) parallel-programming library. To
evaluate these systems, we take two typical end-to-end image analytics
pipelines from astronomy and neuroscience. Each pipeline comes with a
reference implementation in Python provided by the domain
scientists. We then attempt to re-implement them using
the five big data systems and deploy the resulting
implementation on commodity hardware available in the public cloud
to simulate the typical hardware and software setup in
domain scientists' labs.  We then evaluate the resulting
implementations with the following goals in mind:
%
%
%
%
%
\begin{packed_item}
\item Investigate if the given system can be used to implement the
pipelines, and if so how easy is it to do so (\autoref{sec:implementation}).
\item Measure the performance
of the resulting pipelines built on top
of such systems, in terms of execution time when deployed on a cluster
of machines (\autoref{sec:e2e-eval} and \autoref{sec:perOp-eval}).
\item Evaluate the system's ability to scale, both with the number of machines
available in the cluster, and the size of the input data to process (\autoref{sec:e2e-eval}).
\item Assess the tunings, if any, that each system requires to
  correctly and efficiently execute each pipeline (\autoref{sec:optimization}).
\end{packed_item}

Our study shows that, in spite of coming from different domains, the
two real-world use cases have important similarities. Input data takes
the form of multidimensional arrays encoded using domain-specific file
formats (FITS, NIfTI, {\em etc.}).  Data processing involves slicing
along different dimensions, aggregations, stencil (\aka
multidimensional window) operations, spatial joins as well as other
complex transformations expressed in Python.  We find that
all big data systems have important limitations.  SciDB and TensorFlow,
having limited or no support for user-provided Python code,
require rewriting entire use cases in their own languages. Such rewrite
is difficult and sometime impossible due to missing operations. Meanwhile,
optimized implementations of specific operations can
significantly boost performance when available. No system works
directly with scientific image file formats, and all systems require
manual tuning for efficient execution. We could
implement both use cases in their entirety only on Myria and Spark. We
implemented the entire neuroscience use case on Dask also, but found
the tool too difficult to debug for the astronomy use case.
Interestingly, Spark and Myria, which offer data management
capabilities, do so without extra overhead compared with Dask, which
has no such capability. Overall, while performance
and scalability results are promising, we find much room for
improvement in efficiently supporting image analytics at scale.

%% file: systems.tex
\section{Evaluated Systems}
\seclabel{systems}

In this section 
we briefly describe the five evaluated systems 
and their design choices pertinent to image analytics.
The source code of all systems are publicly available.


\textbf{SciDB~\cite{brown:2010}} is a shared-nothing DBMS for storing
and processing multidimensional arrays.  To use SciDB,
users first ingest data into the system, which are
stored as arrays divided into chunks distributed
across nodes in a cluster.  Users then query the stored data using the
Array Query Language (AQL) or Array Functional Language (AFL) through
the provided Python interface. SciDB supports user-defined functions in
C++ and, recently, Python (with the latter executed in a separate
Python process).  In SciDB, query plans are represented as an operator
tree, where operators, including user-defined ones, process data
iteratively one chunk at a time.

\textbf{Spark~\cite{zaharia:12}} is a cluster-computing system.
Spark works on data stored in
HDFS or Amazon S3. 
Spark's data model is centered around the Resilient
Distributed Datasets (RDD) abstraction~\cite{zaharia:12}, where RDDs can 
both reside in memory or on disk distributed across nodes in a cluster.
RDDs are akin to relations
partitioned across the cluster and as such Spark's data model is
similar to that of relational systems. Spark offers a SQL interface,
but users can also manipulate RDDs 
using Scala, Java, or Python APIs, with the latter
executed in a separate Python process as Spark is implemented using Scala.
Programs that manipulate RDDs are represented as graphs. When executed,
the Spark scheduler determines which node in the graph can be executed
and starts by serializing the required objects and data to one
of the machines.

\textbf{Myria~\cite{halperin:14,wang:17}} is a shared-nothing DBMS
developed at the University of Washington. Unlike SciDB and Spark,
Myria can both directly process data stored in HDFS/S3 or ingest data into its own internal representation. 
Myria uses the relational data model and PostgreSQL~\cite{postgres} as its node-local storage
subsystem. Users write queries in MyriaL, an imperative-declarative hybrid language, with SQL-like 
declarative query constructs and imperative statements such as loops. 
Besides MyriaL, Myria supports Python
user-defined functions and aggregates that can be included in
queries. Myria query plans are represented as a graph of operators. When executed,
each operator pipelines data without materializing it to disk. To support Python user-defined
functions, Myria supports the blob data type,
which allows users to write queries that directly manipulate NumPy
arrays or other specialized data types by storing them as blobs.



\textbf{Dask~\cite{dask:2016}} is a general-purpose 
parallel computing library for Python. Dask does not provide data persistence.
Like Spark, Dask distributes data and computation across nodes in a cluster
for arbitrary Python programs. Unlike Spark, users do not 
express their computation using specialized data abstractions. Instead,
users describe their computation using standard Python constructs, except
that computation to be distributed in the cluster is explicitly marked
as {\em delayed} using Dask's API. When Dask
encounters such labeled code, it constructs a compute graph of operators, 
where operators are either Python language constructs or function calls.
When the results of such delayed computation is needed (\eg, they are written
to files), Dask's scheduler determines which 
machine to execute the delayed computation, and serializes the required function
and data to that machine before starting its execution. Unless explicitly instructed,
the computed results remain on the machine where the computation took 
place.

\textbf{TensorFlow~\cite{tensorflow:2015}} is a library for numerical
computation from Google. Like Dask, TensorFlow
does not provide data persistence. It provides C++ and Python 
APIs for users to express
operations over N-dimensional tensors. Such operations are
organized into dataflow graphs, where nodes represent computation, and
edges are the flow of data expressed using tensors (which can be serialized to
and from other data structures such as NumPy arrays).  TensorFlow
optimizes these computation graphs and can execute them locally, in a cluster, on GPUs,
and even on mobile devices. 
Similar to the above systems, the master distributes the computation when deployed on
a cluster. The schedule, however, is specified by the programmer.  
Additionally, all data ingest goes through the master and results are
always returned to the master.





%% file: benchmark.tex
\section{Image Analytics Use Cases}
\seclabel{usecases}

%

We use two real-world scientific image analytics use cases from neuroscience
and astronomy as
evaluation benchmarks.



\subsection{Neuroscience}

Many sub-fields of neuroscience use image data to make inferences
about the brain~\cite{Ji2016-yp}. Data sizes have dramatically grown
recently due to an increase in data collection efforts
\cite{abcdstudy,jernigan2016pediatric}.
The use case we focus on analyzes Magnetic Resonance Imaging (MRI)
data in human brains. Specifically, we focus on diffusion MRI
(dMRI), where the directional profile of diffusion can be used to infer the
directions of brain connections. This method has been used to estimate
large-scale brain connectivity, and relate the properties of brain
connections to brain health and cognitive
functions~\cite{Wandell2016-ms}.

\subsubsection{Data}

The input data comes from the Human Connectome Project~\cite{Van_Essen2013-nd}.
We use data from the S900 release, which includes dMRI data from over 900
healthy adult subjects collected between 2012 and 2015.

The dataset contains dMRI measurements obtained at a nominal
spatial resolution of 1.25$\times$1.25$\times$1.25 $mm^3$.
Measurements were repeated 288 times in each subject, with different
gradient directions and diffusion weightings. Each measurement's data,
called a {\it volume} or {\it image volume}, is stored in a
3D (145$\times$145$\times$174) array of floating point
numbers, with one value per three-dimensional pixel (\aka a voxel).

Information about gradient directions and diffusion weightings is captured in
the image metadata and is reflected in the imageID for each image volume. Each
subject's 288 measurements include 18 in which no diffusion weighting was
applied. These volumes are used for calibration of the diffusion-weighted
measurements.

Each subject's data is stored in standard NIfTI-1~\cite{nifti}
image format and contains the 4D data array (\ie, 288 3D image volumes),
totaling 1.4GB in compressed form, which expands to 4.2GB when uncompressed.
We use up to 25 subjects' data (or a little over 100GB) for this use
case.

\begin{figure}[t]
\centering
  \includegraphics[width=\linewidth]{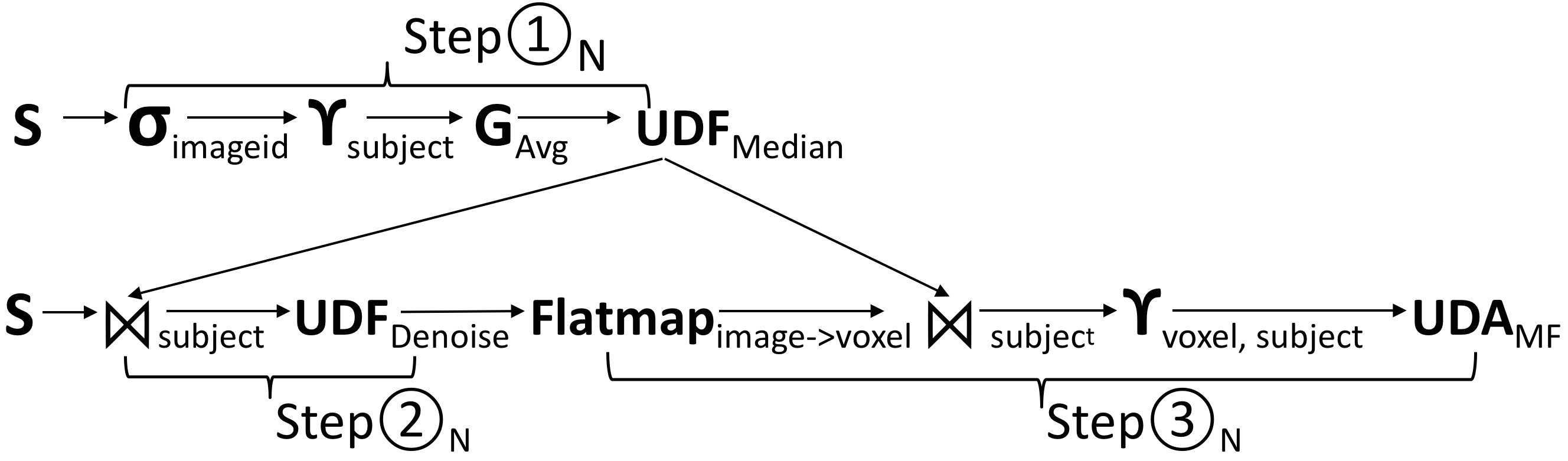}
\caption{Neuroscience use case: Step \circled{1}\textsubscript{N} Segmentation,  Step \circled{2}\textsubscript{N}  Denoising,
and  Step \circled{3}\textsubscript{N}  Model fitting.}
\label{MRIqp}
  \vspace{-0.2in}
\end{figure}

\subsubsection{Processing Pipeline}

Shown in~\autoref{MRIqp},
the benchmark contains three steps from a typical dMRI image analysis pipeline for
each subject.
\neurostep{1} performs volume segmentation
to identify and extract the subset of each image volume that contains the brain
(as opposed to the skull and the background)
\neurostep{2} denoises the extracted image volumes. Finally,
\neurostep{3} fits a physical model of diffusion to each voxel
across all volumes of each subject. We describe each step in detail next.

\begin{figure}[t]
    \centering
    \begin{subfigure}[c]{0.45\linewidth}
  		\includegraphics[width=\linewidth]{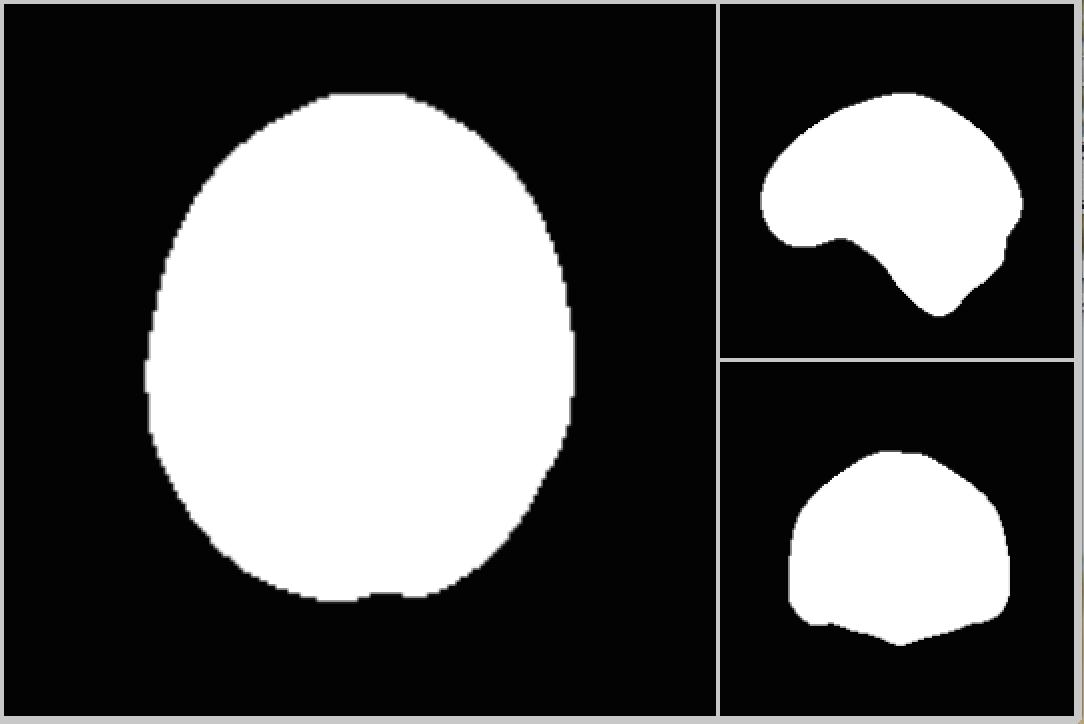}
      \caption{Mask values.}
\label{dmrimask-a}
    \end{subfigure}
    \begin{subfigure}[c]{0.45\linewidth}
    		\includegraphics[width=\linewidth]{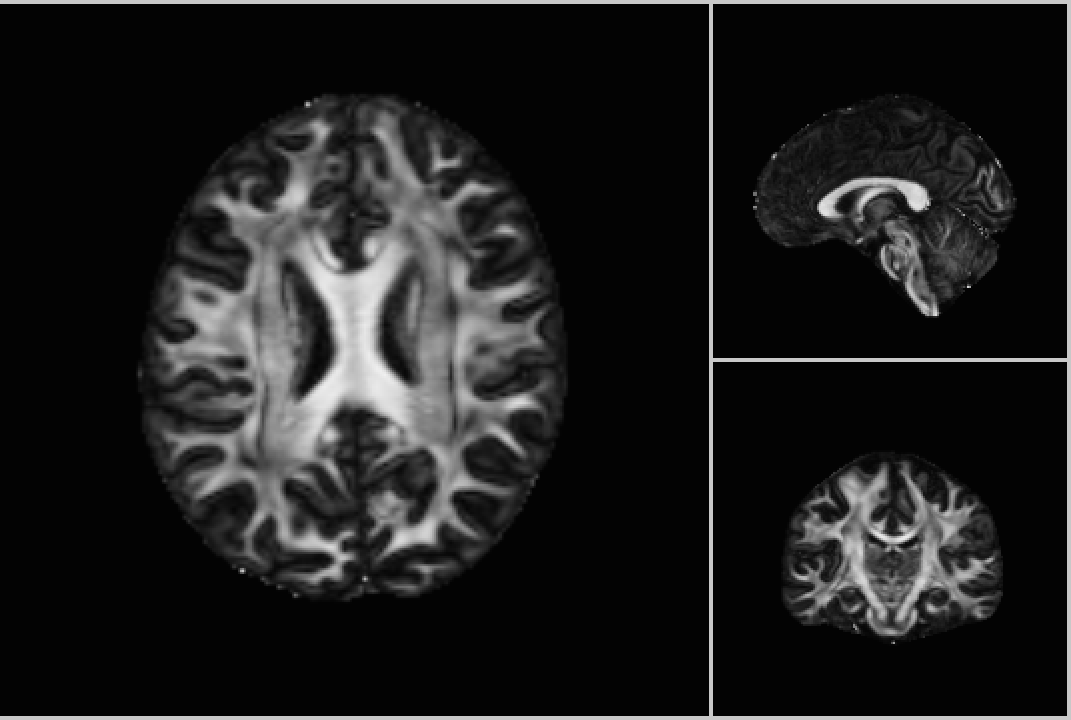}
        \caption{FA values.}
\label{dmrimask-b}
    \end{subfigure}
    \caption{Orthogonal slices of Mask and Fractional Anisotropy values for a single subject.}
         \label{dmrimask}
\vspace{-0.25in}
\end{figure}

\noindent\textbf{Segmentation}: \neurostep{1}
constructs a 3D mask that segments each image volume into
two parts: one with the part to be analyzed, \ie, the
brain, and the other with uninteresting background.
As the brain comprises around two-thirds of the
image volume, using the generated mask to filter out the
background will speed up subsequent steps.
Segmentation proceeds in three sub-steps. First, we
select the subset of volumes with no diffusion weighting applied. These images are used for segmentation
as they have higher signal-to-noise ratio. 
Next, we compute a mean image from the selected volumes by averaging the
value of each voxel. Finally, we apply the Otsu
segmentation algorithm~\cite{Otsu1975-qg} to the mean volume to create a mask
volume per subject.
As an illustration, \figref{dmrimask-a}
shows the orthogonal slices of the binary mask for a single subject.


\noindent\textbf{Denoising}: Denoising is needed to improve image quality and accuracy of the analysis results.
In our pipeline, the denoising step (\neurostep{2})  can be performed on each volume
independently. Denoising operates on a 3D sliding window of voxels using the non-local means
algorithm~\cite{Coupe2008-bx}, where we use the mask from \neurostep{1} to denoise
only parts of the image volume containing the brain.

\noindent\textbf{Model fitting}: Finally, \neurostep{3}
computes a physical model of diffusion. We use the diffusion tensor
model (DTM) for this purpose, which summarizes the directional
diffusion profile within a voxel as a 3D Gaussian
distribution~\cite{Basser1994-hg}. Fitting the DTM is done per voxel
and can be parallelized across voxels. Logically, this step is a
flatmap operation that takes a volume as input and outputs multiple
voxel blocks. All 288 values for each voxel block are then grouped
together before fitting the DTM for each voxel.  Given the 288 values
in a voxel, fitting the model requires estimating a 3$\times$3
variance/covariance matrix (a rank 2 tensor).

The model parameters are summarized as a scalar for each voxel called
Fractional Anistropy (FA) that quantifies diffusivity differences across
different direction. \autoref{dmrimask-b} shows orthogonal slices of the FA values
for a single subject.

Our reference implementation is written in Python and Cython
using Dipy~\cite{Garyfallidis2014-el}
and executes as a single process on one machine.

\subsection{Astronomy}

As discussed in~\autoref{intro}, astronomy surveys are generating an
increasing amount of image data. Our second use case is an abridged
version of the LSST image processing pipeline~\cite{lsstdm}, which
includes analysis routines that astronomers would typically perform on
such data.

\subsubsection{Data}

We use data from
the High cadence Transient Survey~\cite{hits} for this use case,  as data from the LSST survey
is not yet available.
This telescope scans the sky
through repeated \textit{visits} to individual, possibly
overlapping, locations. We use up to 24 visits that cover the same
area of the sky in the evaluation. Each visit is divided into 60
{\it sensor images}, with each consisting of an 80MB 2D image (4000$\times$4072 pixels)
with associated metadata. The total amount of data from all 24 visits is approximately 115GB.

\begin{figure}[t]
\centering
  \includegraphics[width=.65\linewidth]{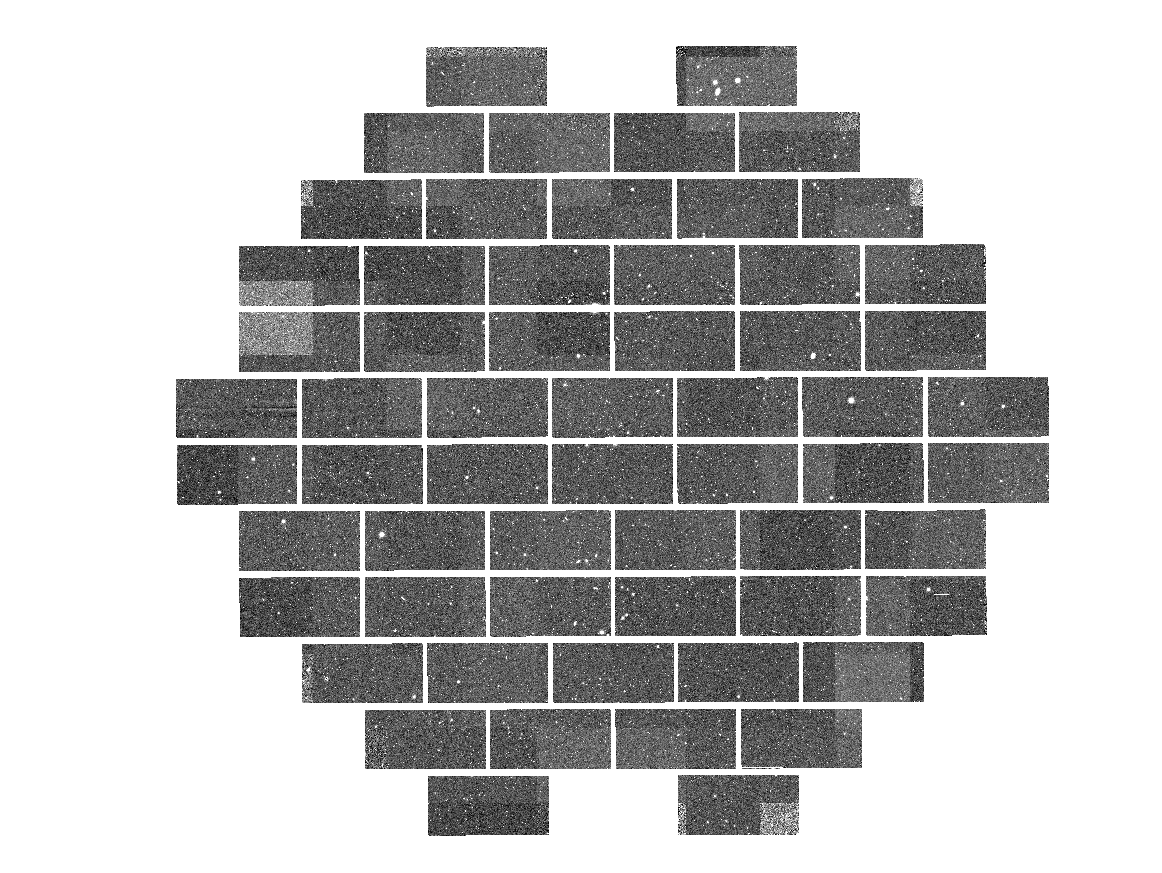}
  \caption{Coadded image of 24 visits to the same place on the sky. Spaces between exposures show sensor boundaries.}
  \label{lsstdata}
  \vspace{-0.2in}
\end{figure}

\autoref{lsstdata} shows the
co-added exposures from 24 visits for one location on the sky. The images are encoded using FITS
format~\cite{fits} with a header and data block. The data block has three 2D
arrays, with each element containing flux, variance, and mask for every
pixel.

\begin{figure}[t]
\centering
  \includegraphics[width=\linewidth]{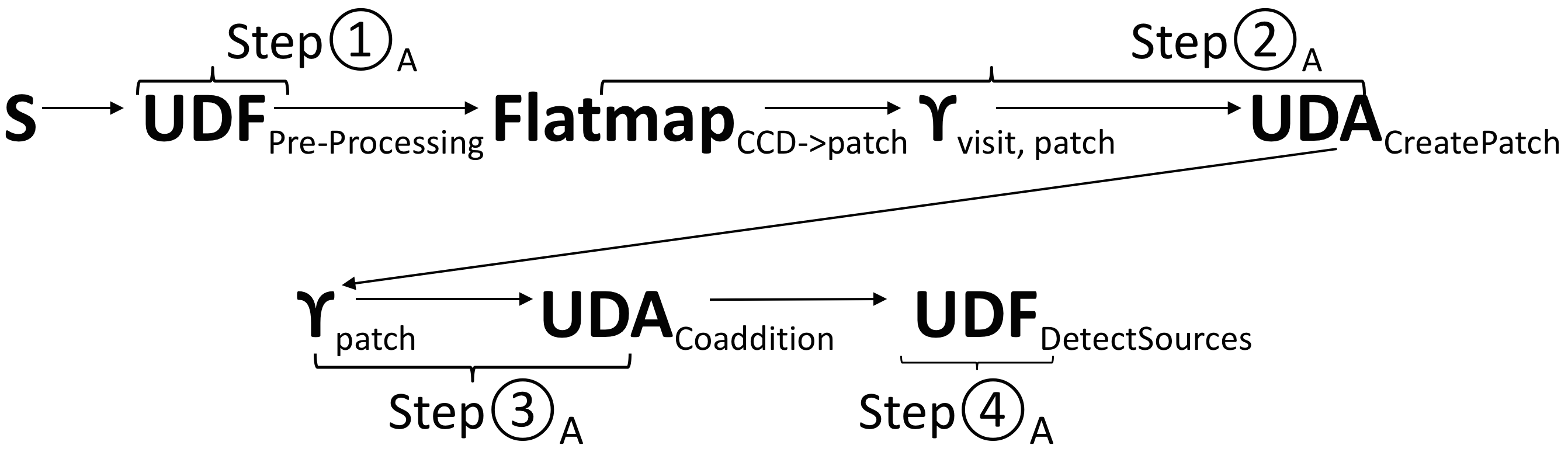}
\caption{Astronomy use case:  \astrostep{1} Pre-Processing,
 \astrostep{2} Patch Creation,
 \astrostep{3} Co-addition, and \astrostep{4}  Source Detection.}
\label{LSSTqp}
\vspace{-0.1in}
\end{figure}

\subsubsection{Processing Pipeline}

Our benchmark contains four steps from the LSST processing pipeline
as shown in \autoref{LSSTqp}:

\noindent\textbf{Pre-Processing}:
We pre-process each input exposure with background estimation and
subtraction, detection and repair of cosmetic defects and cosmic rays, and
aperture corrections for the photometric calibration.  This can be
executed in parallel on each image. The output is called
a \textit{calibrated exposure}.

\noindent\textbf{Patch Creation}:
The analysis partitions the sky into rectangular regions
called {\em patches}. \astrostep{2} maps each calibrated exposure to
the patches that it overlaps. Each exposure can be part of 1 to 6
patches, leading to a logical flatmap operation, which replicates each
exposure once for each overlapping patch.  As pixels from multiple
exposures can contribute to a single patch, this step then groups the
exposures associated with each patch and creates a new exposure object
for each patch in each visit.  The output of this step is an exposure
for each patch.

\noindent\textbf{Co-addition}:  To provide the highest
signal-to-noise ratio (SNR) for the exposures for subsequent
processing, \astrostep{3} groups the exposures associated with the
same patch across different visits and stacks them by summing
up the pixel (or flux) values.  This is called co-addition and the
resulting objects are known as Coadds.  Before summing up the pixel
values, this step performs iterative outlier removal by
computing the mean flux value for each pixel and setting any pixel
that is three standard deviations away from the mean to null. Our
reference implementation performs two such cleaning iterations.

\noindent\textbf{Source Detection}: Finally, \astrostep{4} detects
sources visible in each Coadd generated from \astrostep{3} by
estimating the background and detecting all pixel clusters with flux
values above a given threshold.

Our reference implementation is written in Python, with several
internal functions implemented in C++, utilizing the LSST
stack~\cite{lsst}. While the LSST stack can run on multiple
nodes, the reference is a single node implementation.

%% file: setup.tex
\section{Qualitative Evaluation}
\label{sec:implementation}

We evaluate the five big data systems along two dimensions.  The first
dimension, which we present in this section, is the system's ease of
use, which we measure using lines of code (LoC) needed to implement the use
cases and a qualitative assessment of overall implementation
complexity. We discuss performance and required physical tunings in
\autoref{sec:eval}.

\begin{figure}[t]
\small
\begin{lstlisting}
from scidbpy import connect
sdb = connect('...')
data_sdb = sdb.from_array(data)
data_filtered = data_sdb.compress(sdb.from_array(gtab.b0s_mask), axis=3) # Filter
mean_b0_sdb = data_filtered.mean(index=3)# Mean
\end{lstlisting}
\caption{SciDB implementation of \neurostep{1} in the neuroscience use case. The code
uses the Python interface and AFL to ingest data, filter it using the
\texttt{compress}
function, and compute the
mean along the fourth dimension.}
\figlabel{scidbcode}
  \vspace{-0.2in}
\end{figure}

\subsection{SciDB}
\textbf{Implementation:} SciDB is designed for array analytics to be
implemented in AQL or AFL, optionally on top of SciDB's Python API as
we illustrate in \figref{scidbcode}. SciDB, however, lacks critical
functions including high-dimensional convolutions (\eg, \neurostep{2},
\neurostep{3}, \astrostep{4}), which makes the reimplementation of the
use cases highly nontrivial.  We nevertheless were able to 
rewrite and evaluate two specific operations in SciDB, namely,
\neurostep{1} and \astrostep{3}. SciDB recently released an interface called \texttt{stream()}, which
allows SciDB to pass its array data to an external process 
after converting such data to 
Tab-Separated Values (TSV).  We use this interface to implement
\neurostep{2}.




We implemented two strategies to ingest
ingest neuroscience use case's NIfTI files into SciDB: SciDB-py's built-in API (\ie,
\texttt{from\_array}), and SciDB's accelerated IO library (\ie,
\texttt{aio\_input}).  For the former, we first convert NIfTI
files to NumPy arrays using the NiBabel package~\cite{nibabel} and
import them into SciDB using \texttt{from\_array()}.  For the latter,
we first convert the NIfTI files into Comma-Separated Value (CSV)
files that we then load into SciDB using the \textit{aio\_input}
function.  We use the latter technique for the FITS files from
the astronomy use case.


The SciDB implementation of the neuroscience use case took
155 LoC. \autoref{tab:loc} shows the detailed breakdown by  
operations.  Co-addtion (\astrostep{3}) is expressed in 180 LoC of AQL,
along with 85 LoC Python code for ingesting FITS files into SciDB.

\begin{table}[t]
\centering
\scriptsize
\begin{tabular}{llllll}
\toprule
& \textbf{Dask} & \textbf{SciDB} & \textbf{Spark} & \textbf{Myria} & \textbf{TensorFlow}\footnotemark\ \\
 \midrule

\textbf {Neuroscience} \\
       \textbf{Re-used Reference} & 30 & 3 & 32 & 35 & 0 \\
       \textbf{Data Ingest}   & 33 & 60 & 8 & 5 & 15 \\
       \textbf{Segmentation}     & 25 & 40 & 34 & 10 & 121 \\
       \textbf{Denoising}        & 19 & 52 & 1 & 3 & 128 \\
       \textbf{Model Fit.}     & 11 & NA & 39 & 15 & NA \\
\midrule
\textbf{Astronomy} \\
      \textbf{Re-used Reference}              & X & NA & 212 & 225 & NA \\
      \textbf{Data Ingest}   & X & 85 & 12 & 5 & NA \\
      \textbf{Pre-proc.}      & X & X & 1 & 4 & NA \\
      \textbf{Patch Creation}      & X & X & 4 & 9 & NA \\
      \textbf{Co-Addition}      & X & 180 & 2 & 5 & NA \\
      \textbf{Source Detection} & X & NA & 7 & 2 & NA \\
\bottomrule
\end{tabular}
\caption[Legend]{Lines of code for each implementation. NA not applicable, X not possible to implement.
}
\label{tab:loc}
\vspace{-0.2in}
\end{table}
\footnotetext{\small{Reported LoC 
for each step in TensorFlow contains 64 LoC that are used for all steps.
Reported LoC for segmentation in TensorFlow are only for the mean and filtering computation.}}

\textbf{Qualitative Assessment:} It was challenging to rewrite
the use cases entirely in AQL/AFL.  The recent
\texttt{stream()} interface makes it possible to execute legacy Python
code, but assumes that TSV can be easily digested by the external
process, which required us to convert between TSV and FITS. An
alternate approach would have been to replace FITS handlers with TSV
handlers in the LSST stack, which might have been more efficient but
would definitely be more difficult.


\subsection{Spark}

\textbf{Implementation:} We use Spark's Python API to implement both
use cases. Our implementation transforms the data into Spark's
pair RDDs, which are parallel collections of key-value pair records.
In each RDD, the key attribute is an identifier for an image fragment
and the value is the Numpy array with the image data. Our
implementation then uses the predefined Spark operations ({\tt map}, {\tt flatmap},
{\tt groupby}) to split and regroup image data following the plan from
\autoref{MRIqp}. To avoid joins, we make the mask a
  broadcast variable, which gets automatically replicated on all
  workers. We use the Python functions from the reference
implementation to perform the actual computations on the values. These
functions are passed as arguments, \aka, lambdas, to Spark's map,
flatmap, and groupby operations.  To ingest data in the neuroscience usecase, we first convert
the NIfTI files into NumPy arrays that we stage on Amazon
S3.FITS files staged in s3 as they are. During pipeline execution, we read the Amazon S3 data directly
into worker memory.  \figref{sparkcode} shows an abridged version of
the code for the neuroscience use case.  We implement the astronomy
use case similarly.

\begin{figure}[t]
\small
\begin{lstlisting}
modelsRDD = imgRDD
  .map(lambda x:denoise(x,mask))
  .flatMap(lambda x: repart(x, mask))
  .groupBy(lambda x: (x[0][0],x[0][1]))
  .map(regroup)
  .map(fitmodel)
\end{lstlisting}
\caption{Spark code showing Denoising and Model fitting in the neuroscience use case.
Code for the Python functions is not shown.}
\figlabel{sparkcode}
\vspace{-0.1in}
\end{figure}

\textbf{Qualitative Assessment:} Spark's ability to execute
user-provided Python code over its RDD collections and its
support for arbitrary python objects as keys made the implementation
for both use cases straightforward. We could reuse 
the reference Python code, with fewer than 85 additional LoC
for the neuroscience use case, and fewer than 30 additional LoC
for the astronomy one.  To ensure efficient execution,
however, the Spark implementations of both use cases required tuning
the degree of parallelism and locations for data caching as we
will describe in~\autoref{sec:optimization}.


\subsection{Myria}

\textbf{Implementation:} Myria's use case implementation is similar to
Spark's: We specify the overall pipeline in MyriaL, but call Python
UDFs and UDAs for all core image processing operations.  In the Myria
implementation of the neuroscience use case, we execute a query
to compute the mask, which we broadcast across the cluster. A second
query then computes the rest of the pipeline starting from a broadcast join
between the data and the mask. Also similar to Spark, we read the
NumPy version of the input data directly from S3. In the
neuroscience use case, we ingest the input data into an {\tt Image}
relation, with each tuple consisting of subject ID, image ID and image
volume.  The image volume, containing a serialized NumPy array, 
is stored using the Myria blob data type.
\autoref{myriacode} shows the code snippet for
denoising the image volumes in the neuroscience use
case. \lstref{connect} to~\lstref{registerUDF} connect to Myria and
register the denoise UDF.  \lstref{myrialquery} then executes the
query to join the {\tt Images} relation with the {\tt Mask} relation and denoise
each image volume.

\begin{figure}[t]
\small
\begin{lstlisting}
conn = MyriaConnection(url="...")@\lstlabel{connect}@
conn.create_function("Denoise", Denoise) @\lstlabel{registerUDF}@
query = MyriaQuery.submit("""  @\lstlabel{myrialquery}@
  T1 = SCAN(Images);
  T2 = SCAN(Mask);
  Joined = [SELECT T1.subjId, T1.imgId, T1.img, T2.mask
             FROM T1, T2
             WHERE T1.subjId = T2.subjId];
  Denoised = [FROM Joined EMIT
               PYUDF(Denoise, T1.img, T1.mask) as
               img, T1.subjId, T1.imgId]; """ )
\end{lstlisting}
\caption{Myria code
  showing Python UDF registration and
  execution in the denoising step of the neuroscience use case.}
\figlabel{myriacode}
\vspace{-0.25in}
\end{figure}

We implement the astronomy use case similarly using MyriaL to
specify the overall pipeline and code from the reference
implementation as UDFs/UDAs for image processing.


\textbf{Qualitative Assessment:} Similar to Spark, our Myria
implementation leverages much of the reference Python code, making it
easy to implement both use cases. \autoref{tab:loc} shows that only 2
to 15 extra LoC in MyriaL were necessary to implement each operation.
Myria required a small amount of tuning
for performance and memory management as we describe in
\autoref{sec:optimization}.

\subsection{Dask}
\seclabel{daskImplementation}

\textbf{Implementation:} As described in~\secref{systems}, users specify their computation to Dask
by building compute graphs similar to Spark.
There are two major differences, however.
First, we do not need to reimplement the computation using the RDD abstraction
and can construct graphs directly on top of Python data structures.
On the other hand, we need to explicitly inform Dask when the constructed
compute graphs should be executed, \eg, when the values generated from the graphs
are needed for subsequent computation.

As an illustration of constructing compute graphs,
\figref{daskCode} shows a code fragment from
\neurostep{1} of the neuroscience use case.
We first construct a compute graph that downloads and filters each subject's data
on~\lstref{download}.
Note the use of {\tt delayed} to construct a compute graph
by postponing computation, and specifying that {\tt downloadAndFilter}
is to be called on each subject separately.
At this point, only the compute graph is built; data has not been
downloaded or filtered.


Next, on~\lstref{computeVols} we request Dask to evaluate the compute graph
via the call to {\tt result} to compute the number of volumes in each subject's dataset.
Calling {\tt result} introduces a {\em barrier}, where
the Dask scheduler determines how to distribute the computation to
the threads spawn on the worker machines, executes the graph,
and blocks until {\tt result} returns after
the {\tt numVols} has been evaluated. Since we
constructed the graph such that {\tt downloadAndFilter} is called on
individual subjects, Dask will parallelize the computation
across the worker machines and adjust each machine's load dynamically.

We then build a new compute graph to compute the average image of the
volumes by calling {\tt mean}. We would like this computation to be
parallelized across blocks of voxels, as indicated by
the iterator construct on~\lstref{means}.  Next, the individual
averaged volumes are reassembled on~\lstref{reassemble}, and calling
{\tt median\_otsu} on~\lstref{mask} computes the mask.

\begin{figure} [t]
\small
\begin{lstlisting}[language=python, escapechar=@, numbers=left,
                   numbersep=3pt,numberstyle=\tiny\color{gray}]
  for id in subjectIds:
    data[id].vols = delayed(downloadAndFilter)(id) @\lstlabel{download}@

  for id in subjectIds: # barrier
    data[id].numVols = len(data[id].vols.result())  @\lstlabel{computeVols}@

  for id in subjectIds:
    means = [delayed(mean)(block) for block in
             partitionVoxels(data[id].vols)] @\lstlabel{means}@
    means = delayed(reassemble)(means) @\lstlabel{reassemble}@
    mask = delayed(median_otsu)(means) @\lstlabel{mask}@

\end{lstlisting}
\caption{Dask code showing compute graph construction and
  execution in the segmentation step of the neuroscience use case.}
\label{daskCode}
\vspace{-0.25in}
\end{figure}


The rest of the neuroscience use case follows the same programming
paradigm.  We implemented the astronomy use case with the same
approach.  Interestingly, the implementation freezes once deployed on
a cluster and we found it surprisingly difficult to track down the
cause of the problem. Hence, we do not report performance
numbers for the second use case.


\textbf{Qualitative Assessment:} Dask's compute graph construction API
was relatively easy to use, and we reused most of the reference
implementation.  We implemented the neuroscience use case in
approximately 120 LoC.
In implementing the compute graphs, however, we had to
reason about when to insert barriers to evaluate the constructed
graphs.  In addition, we needed to manually specify how data should be
partitioned across different machines for each of the stages to
facilitate parallelism (\eg, by image volume 
or blocks of voxels,
as specified on \lstref{means} in \autoref{daskCode}). Choosing
different parallelism strategies impacts the correctness and performance
of the implementation. Beyond that, the Dask scheduler did well in
distributing tasks across machines based on estimating data transfer
and computation costs.


\subsection{TensorFlow}

\textbf{Implementation:} As TensorFlow's API operates on tensors, we
need to fully rewrite the reference use case implementation using the
API.  Given the complexity of the use cases, we only implement the
neuroscience one. Additionally, we implement a somewhat simplified
version of the final mask generation operation in \neurostep{1}. We
further rewrite \neurostep{2} using convolutions, but without
filtering with the mask as TensorFlow does not support element-wise data
assignment.


In TensorFlow, the master node handles data distribution: it converts
the input data to tensors, and distributes it to the worker nodes.  In
our implementation, the master reads the data directly from Amazon S3.

TensorFlow's support for distributed computation is currently limited.
The developer must manually map computation and data to each worker
as TensorFlow does not provide
automatic static or dynamic work assignment.  Although the entire use
case could be implemented in one graph,  size limitation
necessitates multiple graphs as each compute graph must be smaller than
2GB when serialized.  Given this limitation, we implement the use case
in steps: we build a new compute graph for each step (as shown
in~\figref{MRIqp}) of the use case. We distribute the data for each
step to the workers and execute the step. We add a global barrier to
wait for all workers to return their results before proceeding, with
the master node converts between NumPy arrays and tensors
as needed. \autoref{fig:tensorflowcode} shows the code for the main
part of the mean computation in \neurostep{1}. The first loop assigns
the input data with shape \texttt{sh} (\lstref{tfDataAssign}) and
the associated code (\lstref{tfWorkAssign}) to each worker. Then, we
process the data in batches of size equal to the number of available
workers on~\lstref{tfExecute}.

\begin{figure}[t]
\small
\begin{lstlisting}[escapechar=@]
# steps contains the predefined mapping
# from data to workernodes
pl_inputs = []
work = []
# The first for loop defines the graph
for i_worker in range(len(steps[0])):
  with tf.device(steps[0][i_worker]):
    pl_inputs.append(tf.placeholder(shape=sh))  @\lstlabel{tfDataAssign}@
    work.append(tf.reduce_mean(pl_inputs[-1]))  @\lstlabel{tfWorkAssign}@
mean_data = []
# Iterate over the predefined steps
for i in range(len(steps)):
  this_zs = zs[i*len(steps[0]):
                i*len(steps[0]) + len(steps[i])]
  # Define the input to be fed into the graph
  zip_d = zip(pl_inputs[:len(steps[i])],
               [part_arrs[z] for z in this_zs])
  # Executes the actual computation - incl. barrier
  mean_data += run(work[:len(steps[i])], zip_d)  @\lstlabel{tfExecute}@
\end{lstlisting}
\caption{TensorFlow code fragment showing compute graph construction
  and execution.}
\label{fig:tensorflowcode}
\vspace{-0.2in}
\end{figure}

\textbf{Qualitative Assessment:} Like SciDB, TensorFlow
requires a complete rewrite of the use case, which takes a significant
amount of time. Similar to Dask, TensorFlow requires that users
manually specify data distribution across machines. The 2GB
limit on graph size further complicates the implementation as we
describe above. Finally, the implementation requires tuning
the degree of parallelism as we will describe in~\autoref{sec:optimization}.



%% file: experiment.tex
\section{Quantitative Evaluation}
\label{sec:eval}

In this section, we evaluate the performance of the implemented
use cases and the system tunings necessary for successful and efficient
execution.  All experiments are performed on the Amazon Web Services cloud
using the r3.2xlarge instance type, with each instance having 8
vCPU\footnote{\small{with Intel Xeon E5-2670 v2 (Ivy
    Bridge) processors.}}, 61GB of Memory, and 160GB SSD storage.

\subsection{End-to-End Performance}
\label{sec:e2e-eval}

\begin{figure*}[t]
    \centering
\scriptsize
    \begin{subfigure}[c]{0.49\textwidth}
       \begin{center}
       \begin{tabular}{lllllll}
          \toprule
          Subjects    & 1 & 2 & 4 & 8 & 12 & 25\\
          \midrule
          Input & 4.1 & 8.4 & 16.8 & 33.6 & 50.4 & 105 \\
          Largest Intermediate & 8.4  & 16.8 & 33.6 & 67.2 & 100.8 & 210 \\
          \bottomrule
       \end{tabular}
       \end{center}
       \vspace{-0.2in}
       \caption{Neuroscience data sizes (GB)}
       \label{mri_datasize}
    \end{subfigure}
    \begin{subfigure}[c]{0.49\textwidth}
      \begin{center}
      \begin{tabular}{llllll}
        \toprule
        Visits    & 2 & 4 & 8 & 12 & 24\\
        \midrule
        Input & 9.6 & 19.2 & 38.4 & 57.6 &115.2 \\
        Largest Intermediate & 24  & 48 & 96 & 144 & 288 \\
        \bottomrule
      \end{tabular}
      \end{center}
       \vspace{-0.2in}
      \caption{Astronomy data sizes (GB)}
      \label{astro_datasize}
    \end{subfigure}
    \begin{subfigure}[c]{0.48\textwidth}
  		\includegraphics[width=\textwidth]{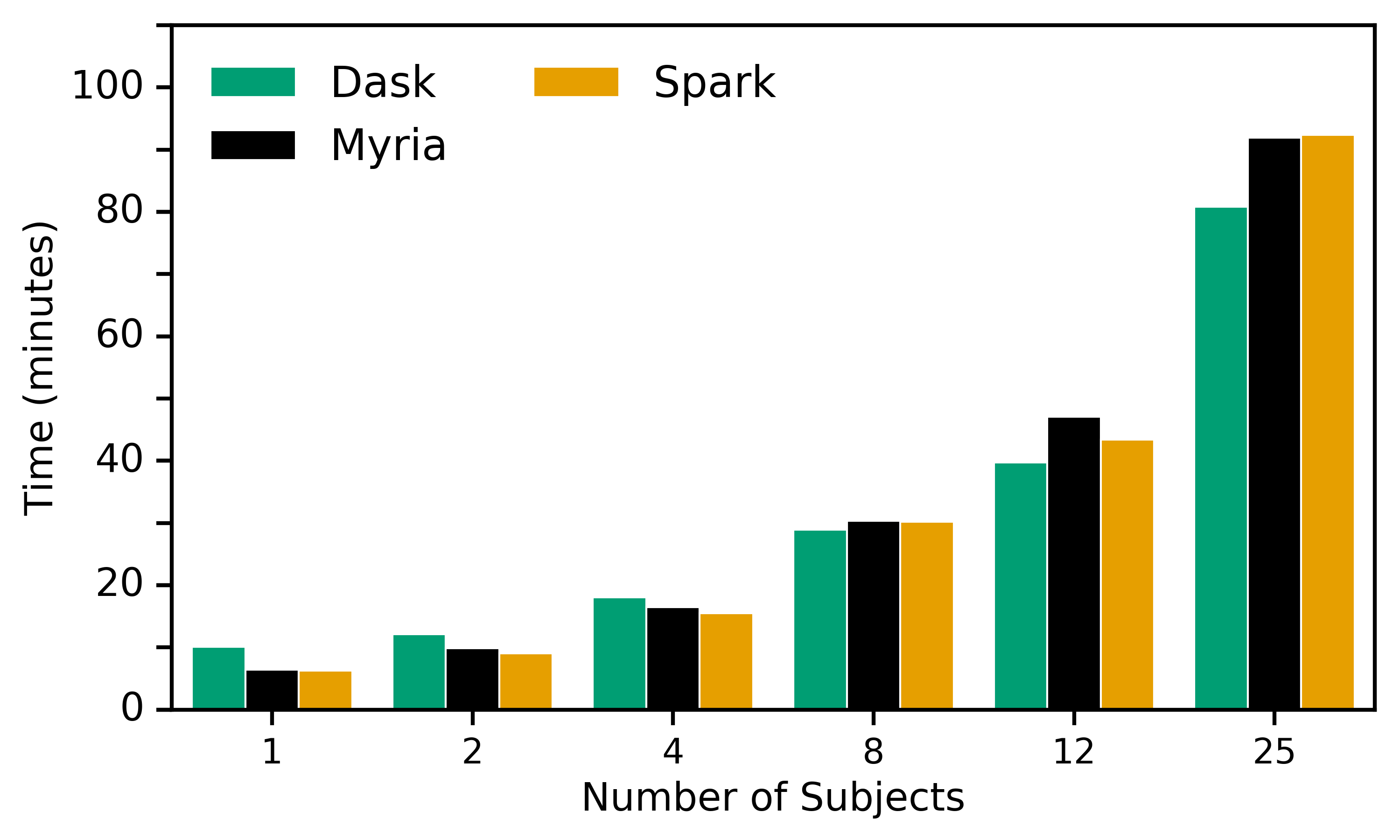}
       \vspace{-0.2in}
  		\caption{Neuroscience: End-to-end runtime with varying data size}
        \label{e2e_data_mri}
  	\end{subfigure}
    \begin{subfigure}[c]{0.48\textwidth}
  		\includegraphics[width=\textwidth]{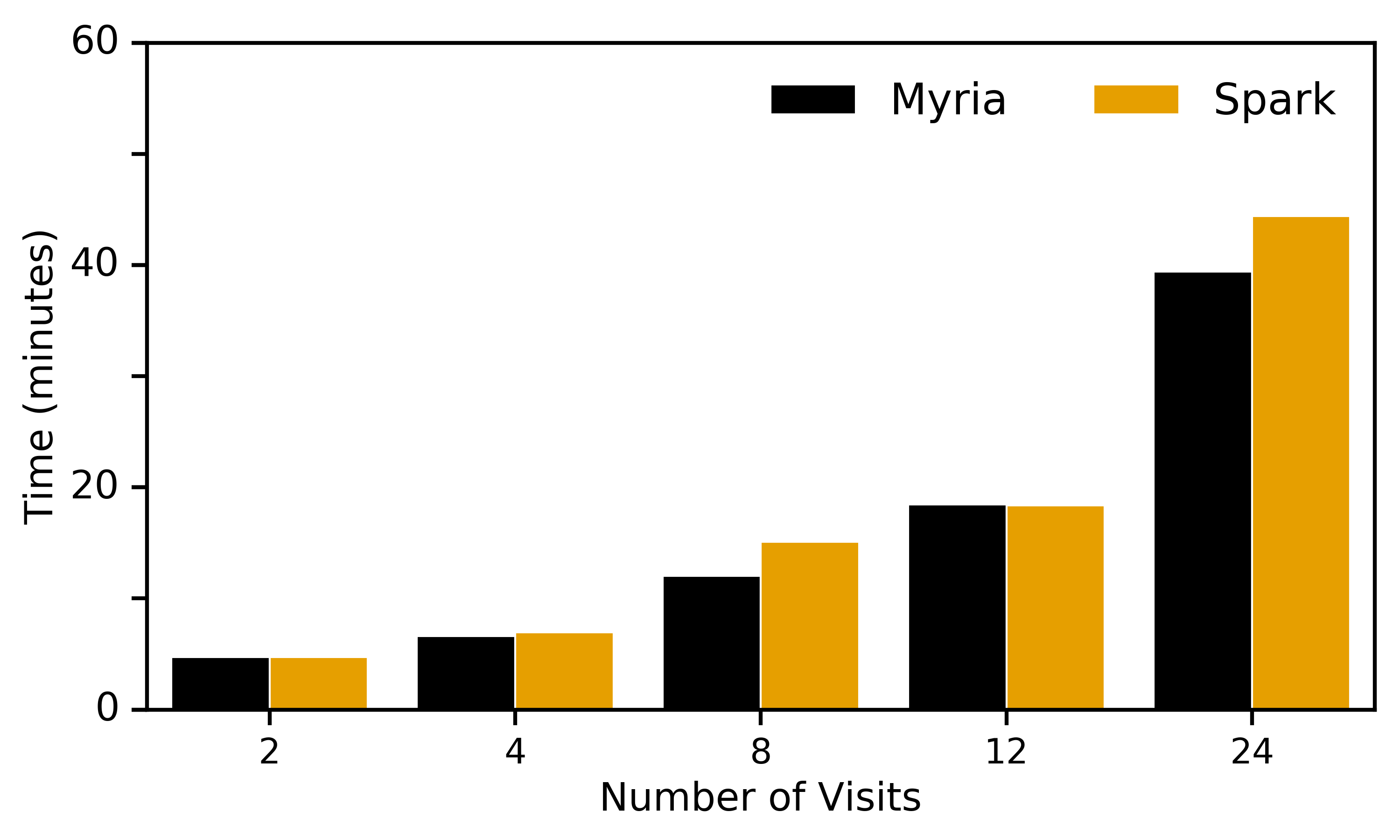}
       \vspace{-0.2in}
  		\caption{Astronomy: End-to-end runtime with varying data size}
      \label{e2e_data_astro}
  	\end{subfigure}
    \begin{subfigure}[c]{0.48\textwidth}
      \begin{center}
      \begin{tabular}{@{} Lllllll@{}}
        \toprule
        Subjects    & 1 & 2 & 4 & 8 & 12 & 25\\
        \midrule
        Dask   & 1 & 0.60 & 0.45 & 0.36 & 0.33 &0.32 \\
        Myria  & 1  & 0.77 & 0.64 & 0.60 & 0.61 & 0.58 \\
        Spark  & 1  & 0.72 & 0.61 & 0.60 & 0.58 & 0.59 \\
        \bottomrule
      \end{tabular}
       \vspace{-0.2in}
      \end{center}
      \caption{Neuroscience: Normalized runtime per subject}
      \label{mri_runtime}
  	\end{subfigure}
    \begin{subfigure}[c]{0.48\textwidth}
      \begin{center}
      \begin{tabular}{@{} Llllll@{}}
        \toprule
        Visits    & 2 & 4 & 8 & 12 & 24\\
        \midrule
        Spark & 1 & 0.73 & 0.79 & 0.64 & 0.78 \\
        Myria & 1  & 0.69 & 0.63 & 0.65 & 0.69 \\
        \bottomrule
      \end{tabular}
       \vspace{-0.2in}
      \end{center}
      \caption{Astronomy: Normalized runtime per visit}
      \label{astro_runtime}
    \end{subfigure}

    \begin{subfigure}[b]{0.45\textwidth}
      \includegraphics[width=\textwidth]{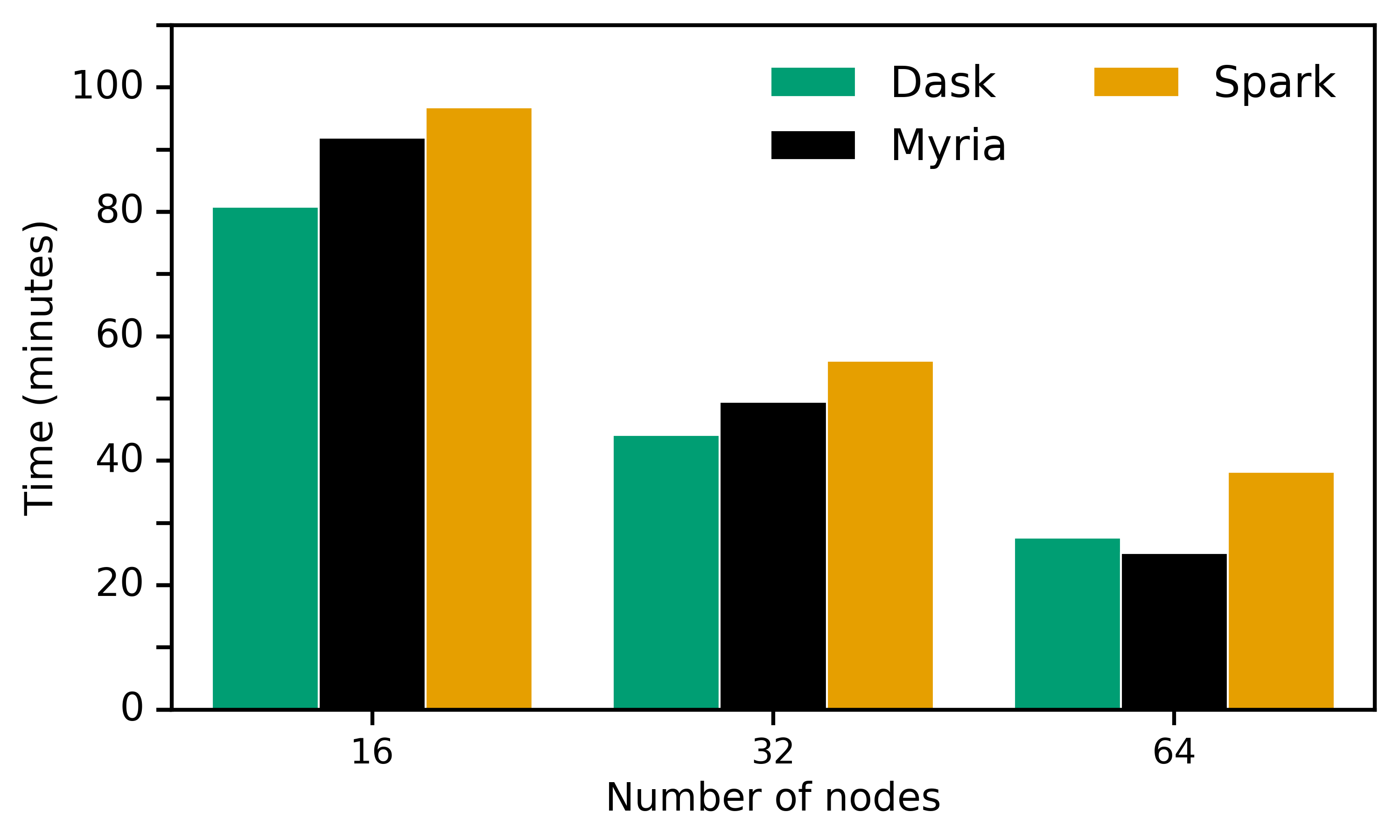}
       \vspace{-0.2in}
      \caption{Neuroscience: End-to-end runtime with varying cluster size}
      \label{e2e_nodes_mri}
    \end{subfigure}
    \begin{subfigure}[b]{0.45\textwidth}
      \includegraphics[width=\textwidth]{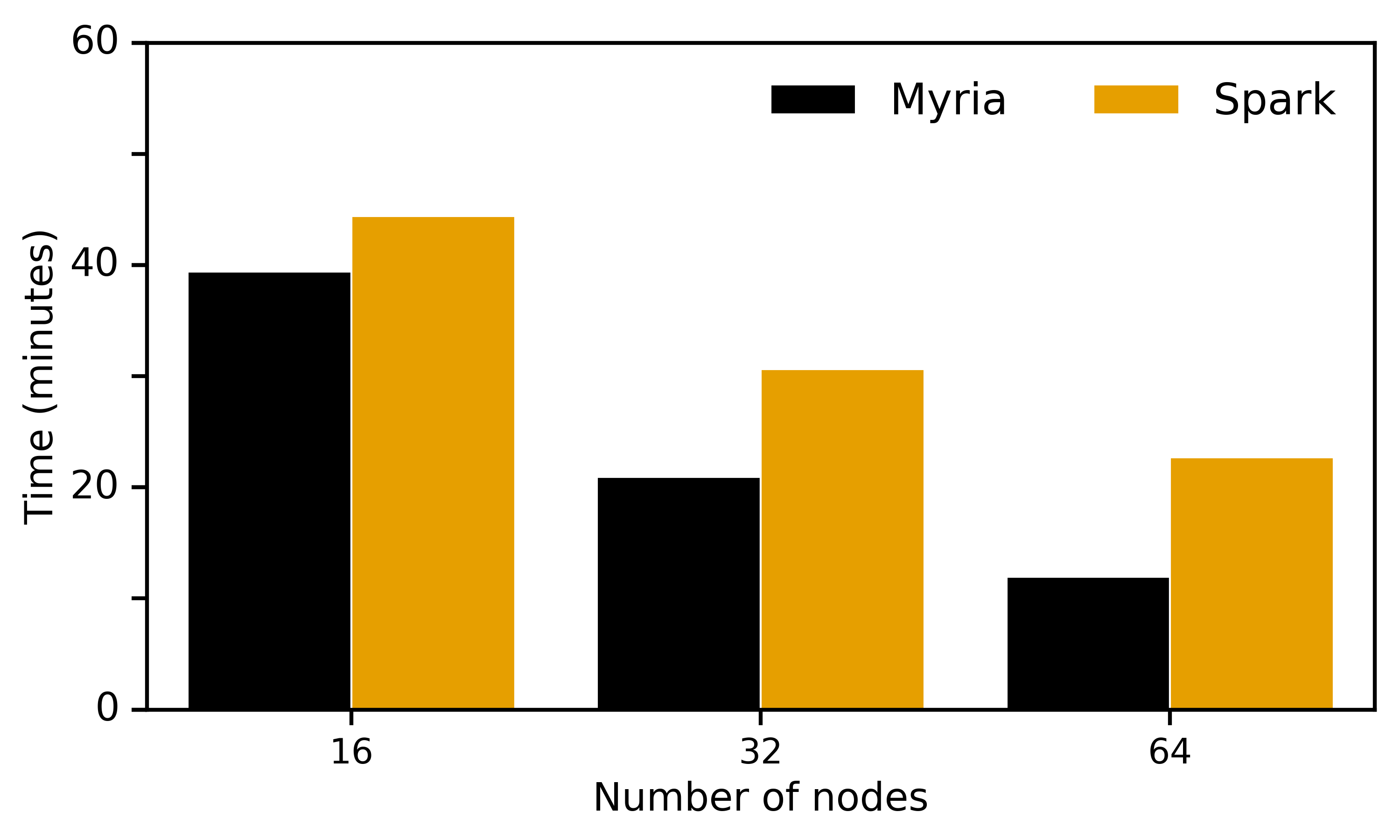}
       \vspace{-0.2in}
      \caption{Astronomy: End-to-end runtime with varying cluster size}
      \label{e2e_nodes_astro}
    \end{subfigure}

    \caption{Overall performance: results for end-to-end experiments for Neuroscience and Astronomy use cases.}
         \label{e2e_results}
         \vspace{-0.2in}
\end{figure*}

We evaluate the performance of running the two use cases
end-to-end. We focus on Dask, Myria, and Spark, the three systems that
we were able to execute one or both benchmarks entirely. We start the
execution with data stored in Amazon S3 and execute all use case
steps. We materialize the final output in worker memories. 
We seek to answer three questions: How does the performance
compare across the three systems? How well do the systems scale as we
grow the input data or the cluster?  What is the penalty, if any, for
executing the use case on top of data processing systems such as Myria
or Spark compared with Dask, a library designed for the distributed
execution of Python code?  To answer these questions, we first fix the
cluster size at 16 nodes and vary the input data size.  For the
neuroscience use case we vary the number of subjects from 1 to 25. The
data for each subject is approximately 4.25GB in size. The input data
thus grows up to a little over 100GB as shown in
\autoref{mri_datasize}.  For the astronomy use case, we vary the
number of visits from 2 to 24. The data for each visit is
approximately 4.8GB in size, for a total of a little over 100GB as
shown in \autoref{astro_datasize}. Because intermediate query results
are larger than the input data, we also show the size of the largest
intermediate relation for each use case. In the second experiment, we
use the largest input data size for each use case and vary the cluster
size from 16 to 64 nodes to measure system speedup.

\autoref{e2e_data_mri} and~\autoref{e2e_data_astro} show the results
as we vary the input data size. All three systems achieve comparable
performance, which is expected as they execute the same Python code on
similarly partitioned data. Interestingly, these results indicate that
there is no significant overhead in using the Myria and Spark data
processing systems compared with simply using Dask. Dask is at best
14\% faster than the other two systems. The faster performance is due
in part to Dask's more efficient pipelining in the context of this
specific use case. In Dask, each subject's data is located on the same
node and processing for the next step can start as soon as the
subject's data has been processed by the preceding step. There are
no dependencies between subjects. Spark and Myria partition data for
every subject across multiple nodes and must thus wait for the
preceding step to output the entire RDD or relation respectively,
and shuffle tuples as needed before proceeding to the next step.
Interestingly, design differences related to data caching and
pipelining during execution (see~\autoref{systems}) do not impact
performance in a visible way between Myria and Spark.  Dask's
performance exhibits a trend somewhat different from the other two
systems. As \autoref{e2e_data_mri} shows, Dask is slower by 60\% for single subject but faster for
larger numbers compared with Spark and Myria.  \autoref{mri_runtime}
and \autoref{astro_runtime} show the runtimes per subject, \ie, the
ratios of each pipeline runtime to that obtained for one subject. As
the data size increases, these ratios drop, indicating that the
systems become more efficient as they amortize start-up costs. Dask's
efficiency increase is most pronounced, indicating that the tool has
the largest start-up overhead.

\autoref{e2e_nodes_mri} and~\autoref{e2e_nodes_astro} show the
pipeline runtimes for all systems as we increase the cluster size and
process the largest, 100GB datasets. All systems show near linear
speedup for both use cases.  Myria achieves almost perfect linear speedup. Dask is better than
Myria on smaller cluster sizes but scheduling
overhead makes Dask less efficient as cluster sizes increase, as the
scheduler attempts to move tasks among different machines via aggressive work
stealing.

Note that we tuned each of the systems to achieve the timings reported above.
We discuss the impact of these tunings in \autoref{sec:optimization}.

\subsection{Individual Step Performance}
\label{sec:perOp-eval}

Next we focus on the performance of a subset of the pipeline
operations that we successfully implemented in TensorFlow and SciDB in
addition to the other three systems above.

\subsubsection{Data Ingest}

The input data for the use cases was staged in Amazon S3. While Myria
and Spark can read data directly from S3, repeated operations on the same data
can benefit from data being cached either in cluster memory (for Spark) or
in local storage across the cluster (for Myria). Other systems must
ingest data before processing it. In this section, we measure the time
it takes to ingest data.

\begin{figure}[t]
	\centering
		\includegraphics[width=0.95\linewidth]{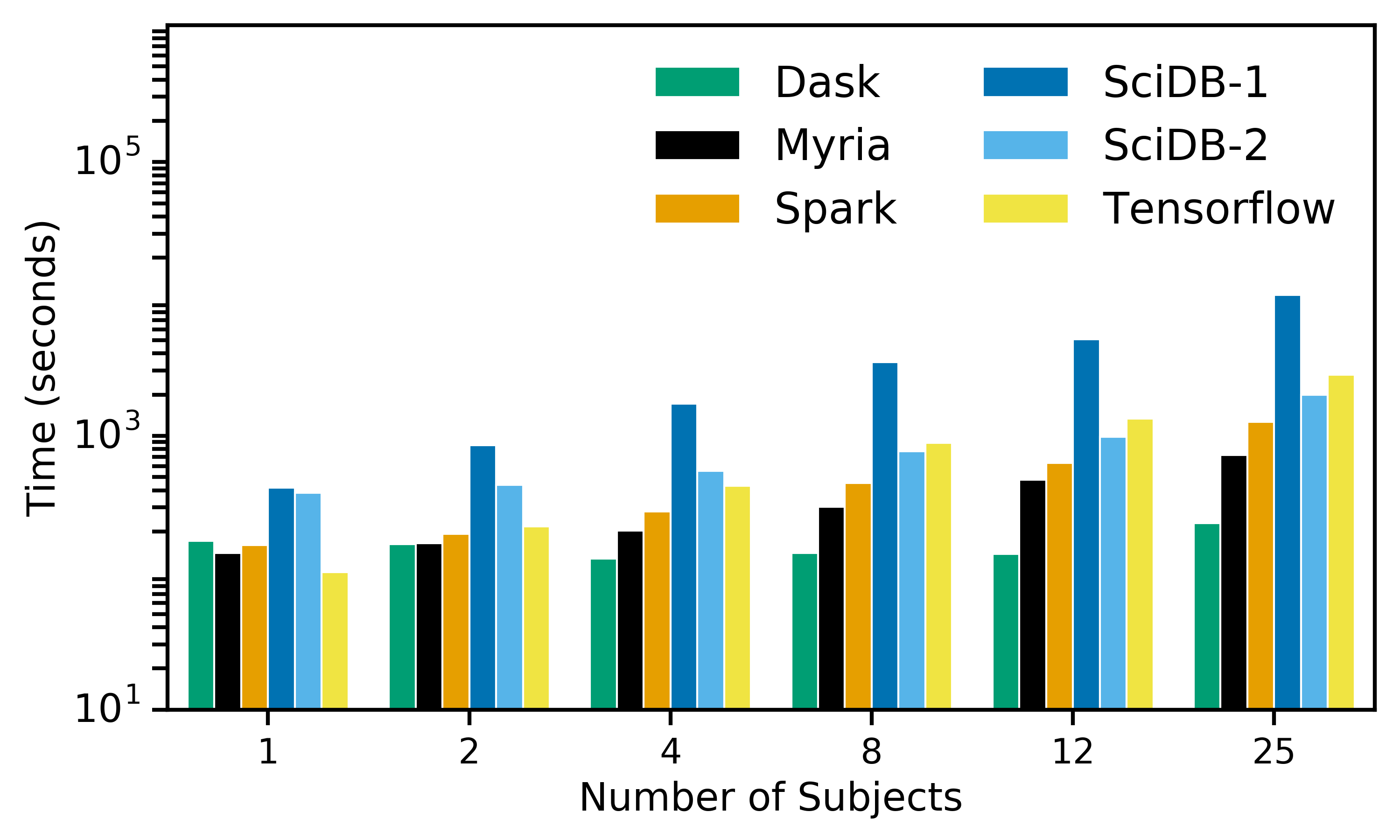}
		\caption{Data ingest times for the neuroscience
                  benchmark on a 16-node cluster. Y-axis uses a log
                  scale.}
	  \label{dataingest}
\vspace{-0.2in}
\end{figure}

For Myria, we measure the time to read data from S3 and store it
internally in the per-node PostgreSQL instances. For Spark, we measure
the time to load data into in-memory RDDs.  For both systems, we first
preprocess the NifTi files into individual image volumes, which we
persist as pickled NumPy files per image in S3 before the ingest, the conversion time is included in the
data ingest time. For Dask and
TensorFlow, we measure the time to load the data from NifTi files into
in-memory NumPy arrays. Finally, for SciDB, we measure the time to
ingest the data from NIfTI files into an internal multidimensional
array spread across the SciDB cluster.  \autoref{dataingest} shows the
results.

As the figure shows, data ingest times vary greatly across systems
(note the log scale on the y axis).  Spark and Myria both download
data in parallel on each of the workers from S3.  Spark's API requires
the name of the S3 bucket and enumerates the data files on the master
node before scheduling the parallel download on the workers, while
Myria can directly work with a csv list of files avoiding overhead and is
faster than Spark for this step even though it writes the data locally
to disk.

For Dask, we explicitly specify the number of subjects to download per
node, since each machine only has enough
memory to fit the data for up to 3 subjects, and Dask's scheduler would
schedule random number of subjects per machine as it does not know how
much data will be downloaded. As a result, Dask's data ingest time remains constant until the
number of subjects exceeds 16, and we assign a subset of machines to download
data for more than one subject. The TensorFlow implementation downloads all data to the master
node and sends partitions to each node in a pipelined fashion, which
is slower than the parallel ingest available in other tools.

We report two sets of timings for SciDB in~\autoref{dataingest}.
SciDB-1 reports the time to ingest NumPy arrays with {\tt from\_array()} interface,
and SciDB-2 reports the time to convert NIfTI to CSV and ingest using the
{\tt aio\_input} library. The {\tt aio\_input()} function is an order of magnitude faster than the pythonAPI based ingest for SciDB and is
 on par with parallel ingest on Spark and Myria. However, the
NIfTI-to-CSV conversion overhead for SciDB is a little larger than the NIfTI-to-NumPy overhead for Spark and Myria, which makes
SciDB ingest slower than Spark and Myria.

\subsubsection{Neuroscience Use Case: Segmentation}


Segmentation is the first step in the neuroscience use case (\ie,
\neurostep{1}). We measure the performance of two operations in this step:
filtering the data to select a subset of the image volumes, and
computing an average image volume for each subject.  \autoref{filter}
and \autoref{mean} show the runtimes for these two operations as we
vary the input data size on the 16-node cluster.

Myria and Dask are the fastest systems on the data filtering
step. Myria pushes the selection down to PostgreSQL, which efficiently
scans the data and returns only the matching records. Dask is also fast
on this operation as all its data is already in memory and the
operation is a simple filter. Spark is an-order of magnitude slower
than Dask, even though data is in memory for both systems. This is due to
Spark's need to serialize Python code and data among machines.
In TensorFlow, the data (tensors) takes the form of 4D arrays. For each subject,
the 4D array represents the 288 3D data volumes. We need to filter on
the volume ID, which is the fourth dimension. TensorFlow,
however, only supports filtering along the first dimension. We thus
need to flatten the 4D array apply the selection, and reshape the array back into a 4D structure. As
reshaping is expensive compared with filtering, TensorFlow
is orders of magnitude slower than the other engines on this operation.

SciDB is slower than some other systems mainly because the internal chunks are
not aligned with the selection.  That is, in addition to simply reading data
chunks, SciDB does more work including extracting subsets out of the chunks and
reconstructing them into the resultant arrays.

\autoref{mean} shows the result for the mean image volume computations.
SciDB is the
fastest for mean computation on the small datasets as it is optimized
for array operations and this computation exercises SciDB's
specialized design. Spark and Myria demonstrate super-linear
scalability for this operation and are comparable with SciDB at larger
scales. This is because, at smaller scales, the number of workers used
during the mean operation is approximately equal to the number of groups, and there
is one group per subject, resulting in low cluster utilization. Dask,
meanwhile, performs slower than the other three engines for small datasets due to
startup overheads and the overhead of aggressive work stealing.
TensorFlow incurs extra cost in converting from image
volume to tensors and is an order of magnitude slower than the other engines.

\subsubsection{Neuroscience Use Case: Denoising}

\autoref{denoise} shows the runtime for the denoising \neurostep{2}.
For this step, the bulk of the processing happens in the user-defined
denoising function. Dask, Myria, Spark, and SciDB all run the same code from
the reference implementation on similarly partitioned data, which
leads to similar overall performance. As in the case of the
end-to-end pipeline, Dask's higher start-up overhead combined with
more efficient subsequent processing leads to slightly worse
performance for smaller data sizes but similar performance as the data
grows. TensorFlow, once again, incurs the overhead of data conversion
between image volumes and tensors. Additionally, in the TensorFlow
implementation, we could not use the mask to reduce the computation
for each data volume as TensorFlow's operations can only be applied to
whole tensors and cannot be masked.
SciDB's \texttt{stream()} interface, although allows us to automatically
process SciDB's array chunks in parallel, performs slightly worse than Myria,
Spark, and Dask.  This is due to the fact that \texttt{stream()} connects
SciDB and external processes only through data in CSV format, and doing so
incurs significant overhead.

\subsubsection{Astronomy Use Case: Co-addition}

\begin{figure*}[!htbp]
	\centering
	\begin{subfigure}[c]{0.45\textwidth}
		\includegraphics[width=\textwidth]{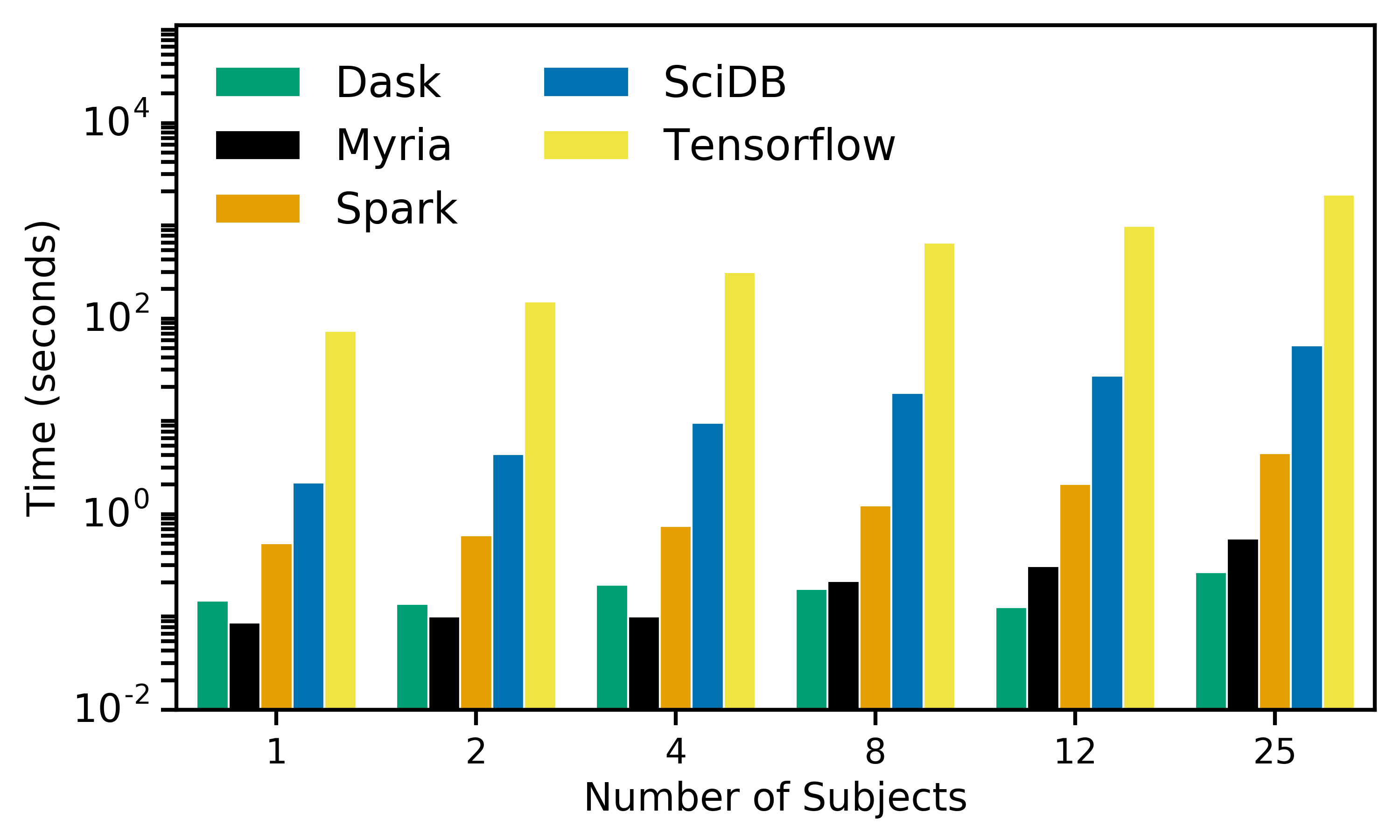}
    \vspace{-0.2in}
		\caption{Filter (Segmentation--Neuroscience use case)}
    \label{filter}
	\end{subfigure}
	\begin{subfigure}[c]{0.45\textwidth}
		\includegraphics[width=\textwidth]{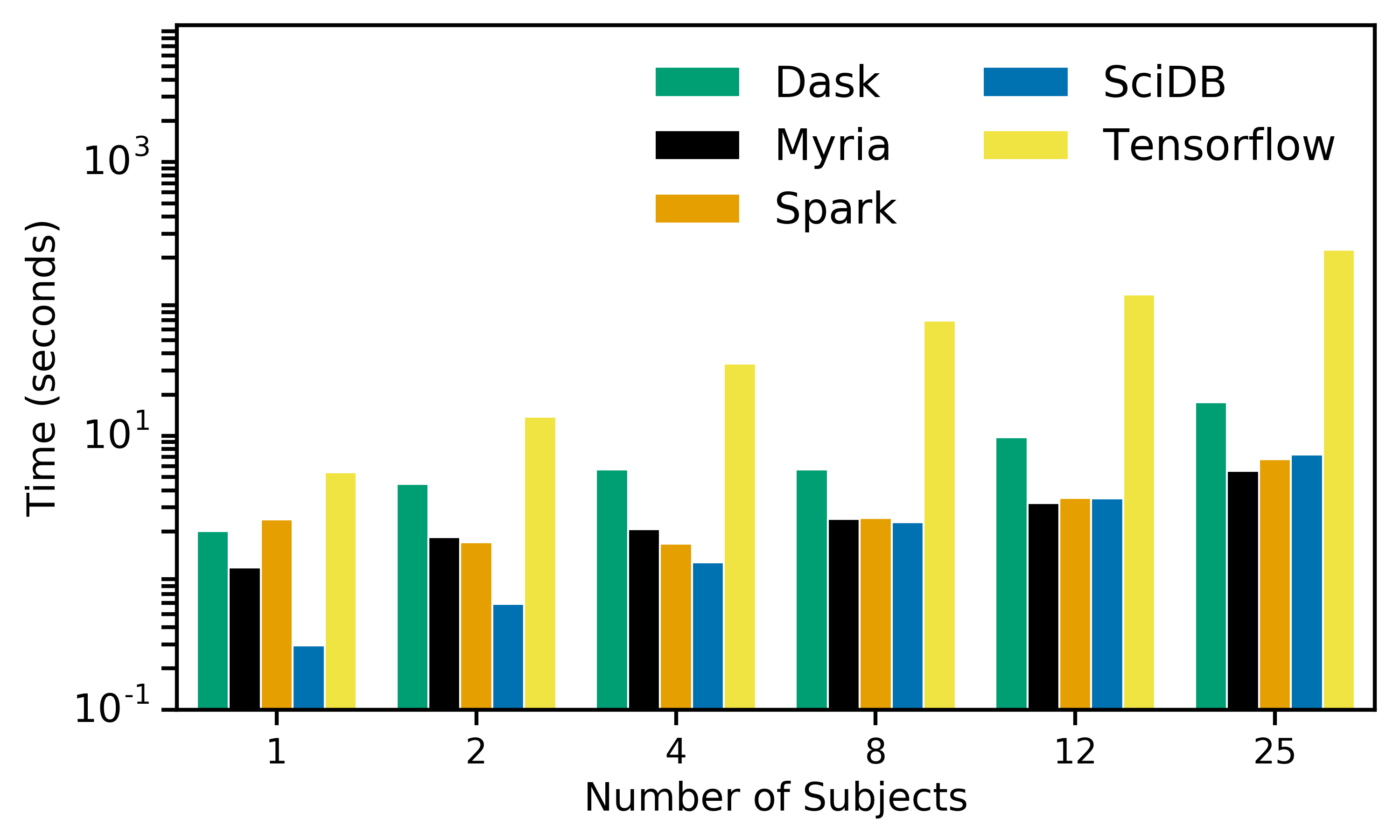}
       \vspace{-0.2in}
		\caption{Mean (Segmentation--Neuroscience use case)}
    \label{mean}
	\end{subfigure}
	\begin{subfigure}[b]{0.45\textwidth}
		\includegraphics[width=\textwidth]{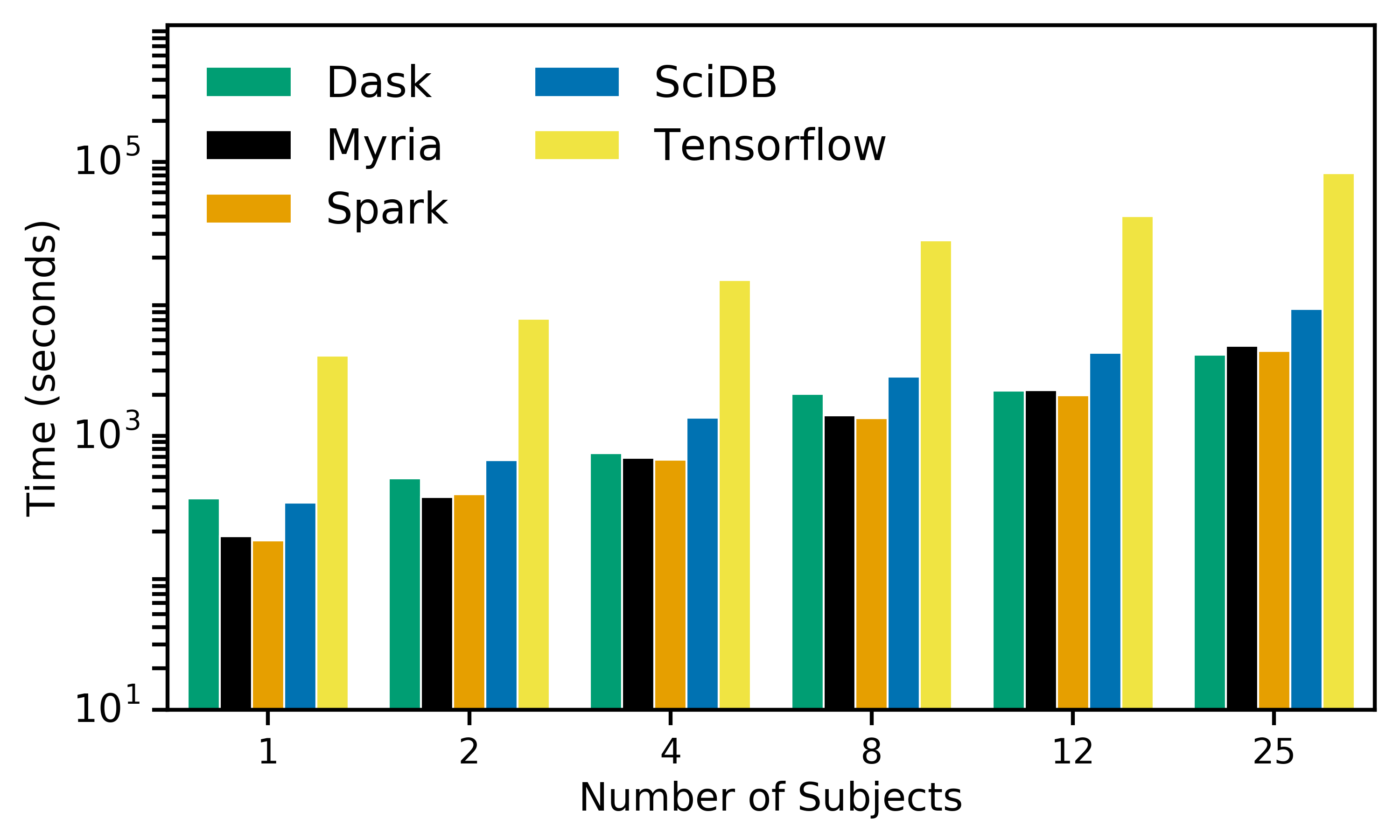}
    \vspace{-0.2in}
		\caption{Denoise--Neuroscience use case}
    \label{denoise}
	\end{subfigure}
	\begin{subfigure}[b]{0.45\textwidth}
		\includegraphics[width=\textwidth]{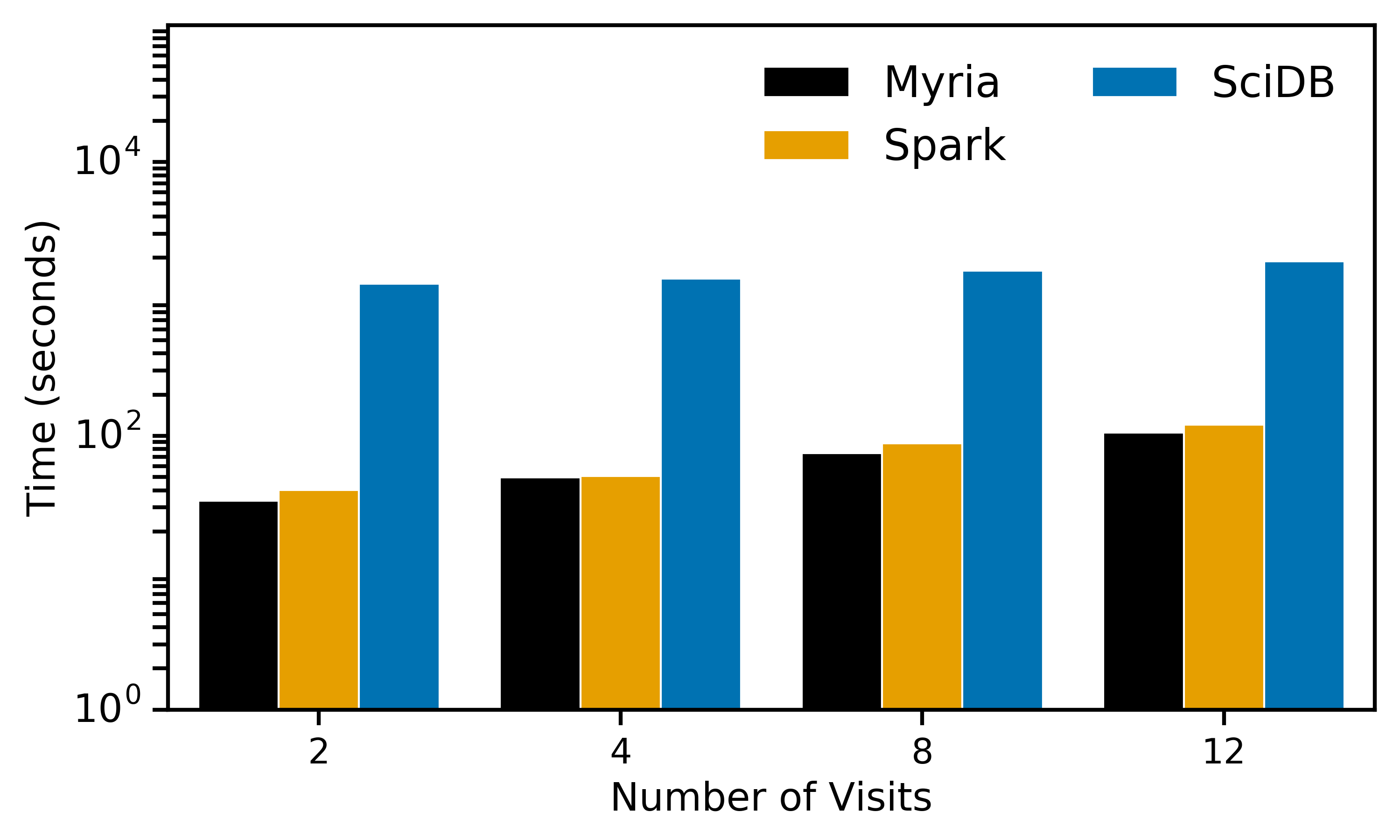}
    \vspace{-0.2in}
		\caption{Coaddition--Astronomy use case}
    \label{coadd}
	\end{subfigure}
  \caption{\textbf{Individual step performance} (log scale on the y-axis).
  Experiments run on 16 nodes with largest dataset.}
  \label{steps}
  \vspace{-0.2in}
\end{figure*}

Finally, \autoref{coadd} shows the runtimes for the coaddition
\astrostep{3}.  This step is interesting because it involves iterative
processing. Once again, Spark and Myria leverage the reference
implementation as user-defined code. The reference implementation performs
iterative processing internally, yielding high performance. For
SciDB, we reimplemented this step entirely in AQL. Furthermore, we use
the official SciDB release, which does not include any optimizations
for iterative processing, resulting in runtimes more than one order of
magnitude slower than those of the other two engines. We observe,
however, that our prior work~\cite{Soroush_ssdbm15} proposed
effective optimizations for iterative processing in array engines. By
extending SciDB with incremental iterative processing, we showed a
6$\times$ improvement in the execution of that same step. With this
optimization, SciDB's performance would be on par with Spark and Myria
for the larger data sizes. Additionally, we could also implement this
step using the new stream interface in SciDB, which should then yield
similar performance to Myria and Spark as it would execute the Python
reference code. Overall, this experiment shows the importance of
efficient iterative processing for real-world image analytics
pipelines.

\subsection{System Tuning}
\label{sec:optimization}

Finally, we evaluate the five systems on the complexity of the tunings
required to achieve high performance.

\subsubsection{Degree of Parallelism}
\label{sec:optimize:parallel}

Degree of parallelism is a key parameter for big data 
systems that depends on three factors: (1) the number of nodes in the
physical cluster; (2) the number of workers that can execute
in parallel on each node; and (3) the size of the data partitions that
can be allocated across workers.  We evaluated the impact of
changing the cluster size in~\autoref{sec:e2e-eval}. In this section,
we evaluate the impact of (2) and (3).

\begin{figure}[t]
\centering
\includegraphics[width=0.8\linewidth]{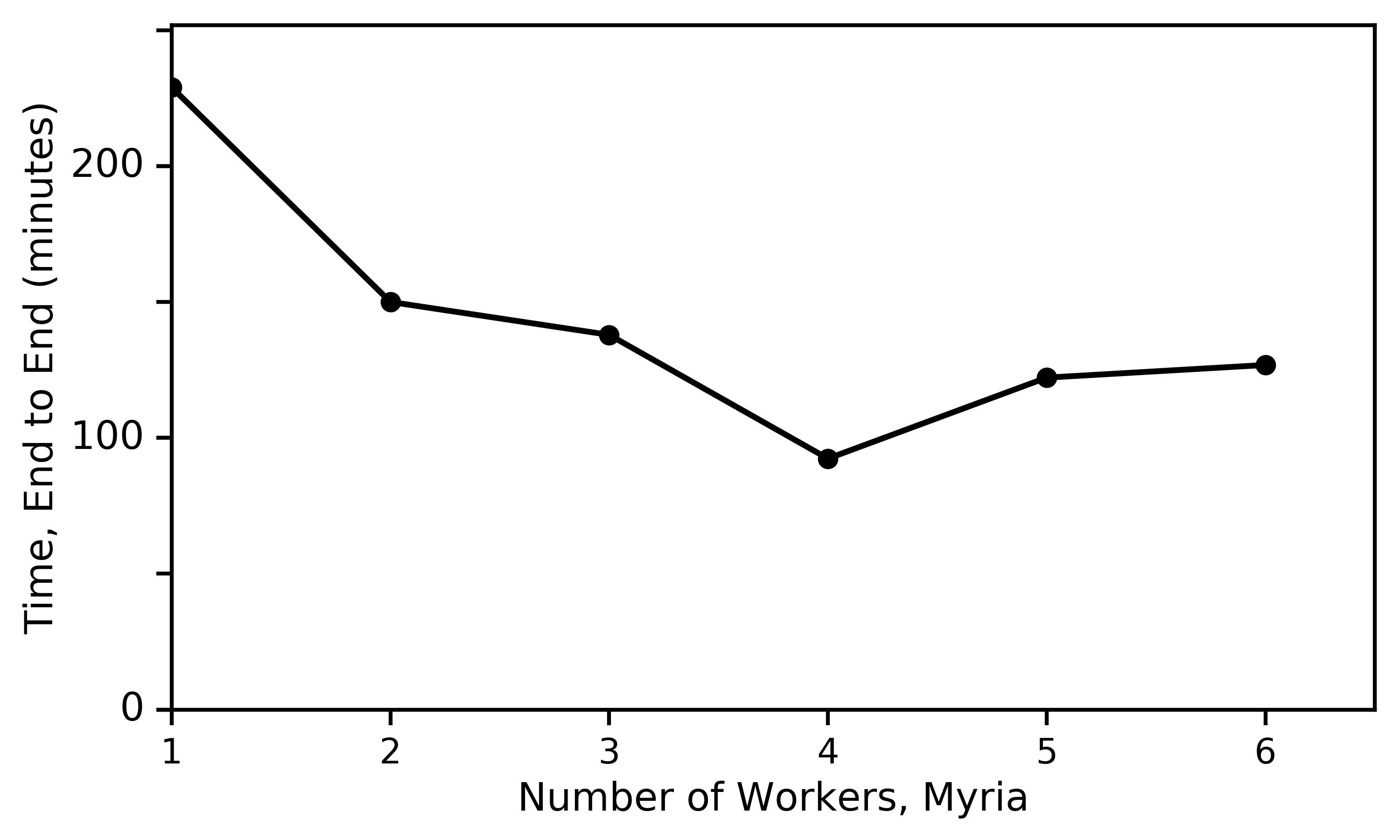}
\caption{Execution time for 25 subjects in the neuroscience use case on Myria for
  different number of workers per node on a 16 node cluster.}
       \label{mri-workers}
\vspace{-0.1in}
\end{figure}

For Myria, given a 16-node cluster, the degree of parallelism is
entirely determined by the number of workers per node. As
in traditional parallel DBMSs, Myria hash-partitions data
across all workers by default.  \autoref{mri-workers} shows
runtimes for different numbers of workers for the
neuroscience use case. A larger number of workers yields a higher
degree of parallelism but workers also compete for physical resources
(memory, CPU, and disk IO). Our manual tuning
found that four workers per node yields the best results.
The same holds for the astronomy
use case (not shown due to space constraints).

\begin{figure}[t]
\centering
\includegraphics[width=0.8\linewidth]{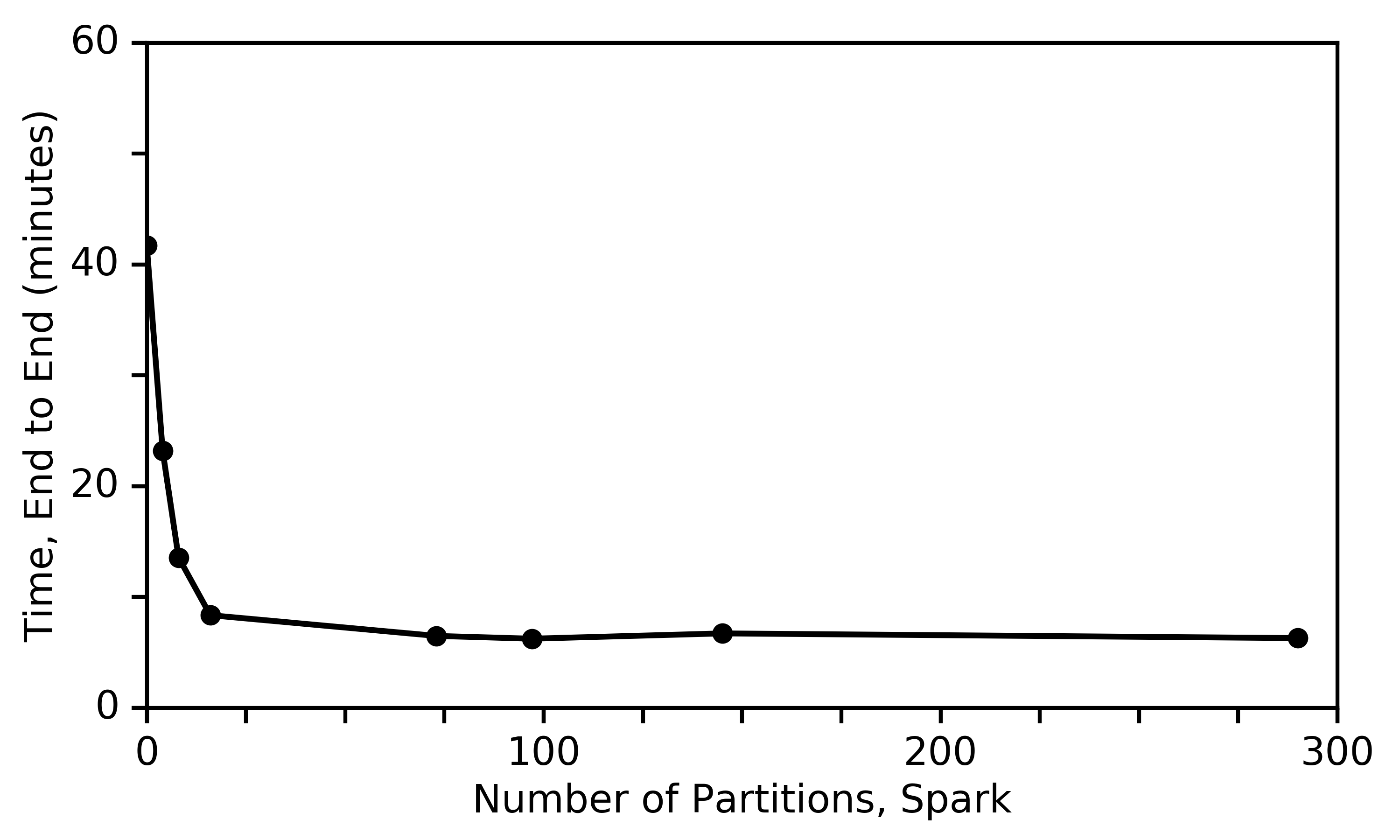}
\caption{Execution time for a single subject in the neuroscience use case on Spark with
  varying number of input data partitions. }
\label{mri-partitions}
\vspace{-0.2in}
\end{figure}

Spark, unlike Myria, creates data partitions (which
correspond to \textit{tasks}), each worker can execute as many tasks in
parallel as available cores. Thus, number of workers per node does not impact degree of
parallelism when number of cores remains the same. The number of data partitions determines the number
of schedulable tasks. \autoref{mri-partitions} shows the query
runtimes for different numbers of input data partitions.  On a 16-node
cluster, the decrease in runtime is dramatic between 1 and 16
partitions, as an increasingly large fraction of the cluster becomes
utilized. The runtime continues to improve until 128 data partitions
which is the total number of slots in the cluster (\ie, 16 nodes $\times$ 8
cores). Increasing the number of partitions from 16 to 97 results in 50\% improvement. Further increases do not improve
performance. Interestingly, if the number
of data partitions is unspecified, Spark creates
a partition for each HDFS block, which typically leads to a small
number of large partitions for the data that we use. For
example, for the neuroscience use case with a single subject, Spark
creates only 4 partitions that results in a highly underutilized
cluster. Overall, Spark and Myria both require tuning to achieve best
performance. Tuning is simpler in Spark, however, because any
sufficiently large number of partitions yields good performance.

\begin{figure}[t]
\centering
\includegraphics[width=0.8\linewidth]{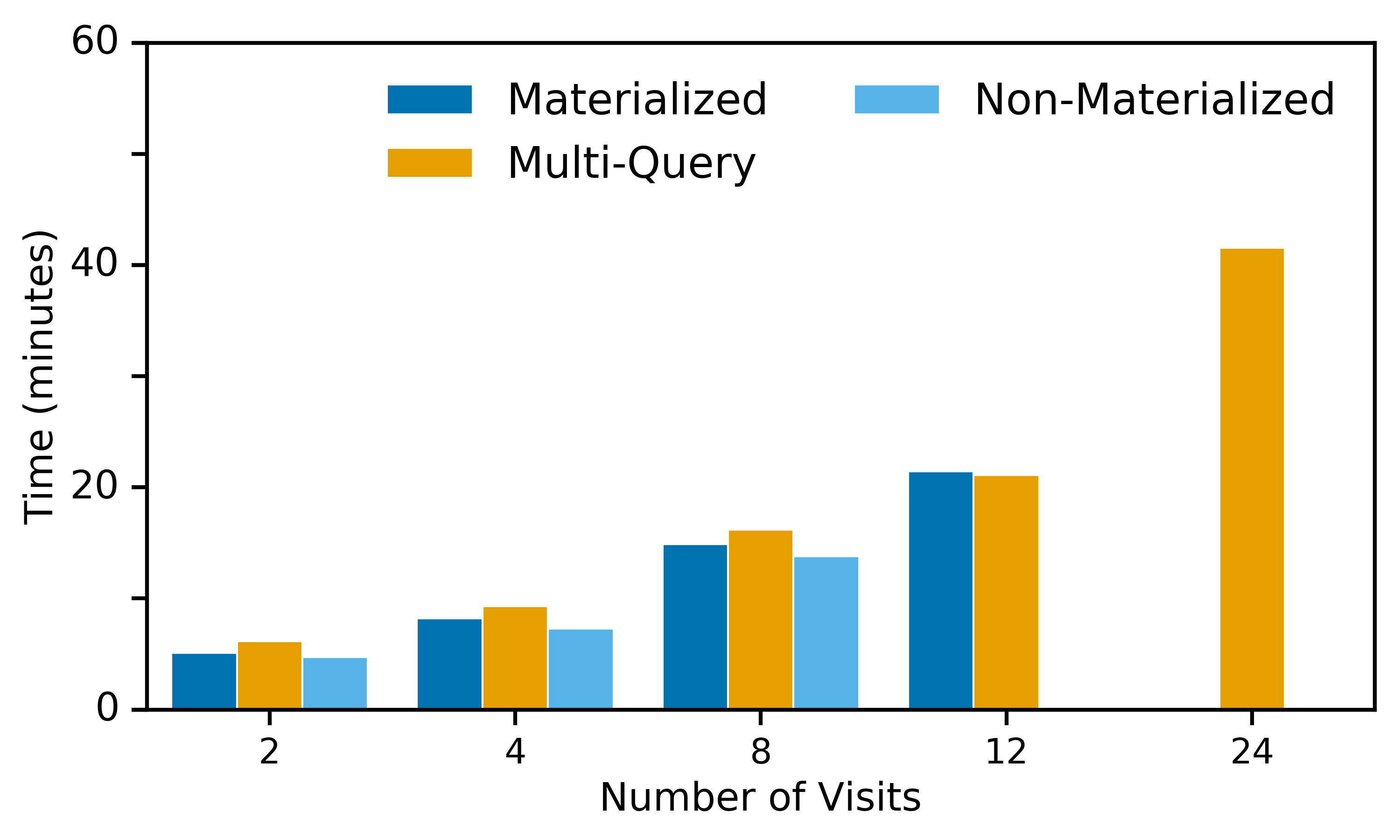}
      \caption{Execution time for the astronomy use case on Myria
        using different memory management techniques.}
      \label{astro-mat}
\vspace{-0.2in}
\end{figure}

In TensorFlow, we executed one process per machine. For most operations, data conversions rather
than other resources were the bottleneck eliminating the need for
additional tuning. For the denoising step, memory was the bottleneck,
which required the assignment of one image volume per physical
machine. For filtering, we experimented with assigning different
numbers of image volumes at a time to different workers on different
machines and found a factor of 2$\times$ difference in total runtime between
different assignments.
Dask does not come with any partitioning capability and we manually
tuned the number of workers and data partitions similar to Spark.
Dask's work stealing automatically load balances across the machines, however, work-stealing did not need any tuning.

In SciDB, it is good practice~\cite{scidb_instance_core}
to run one instance per 1-2 CPU cores.  The chunk size, however, is more difficult to tune,
as we did not find a strong correlation between the overall
performance and common system configurations.  We thus estimated the
optimal chunk size for each operation by running the application with
the same data set but using different chunk sizes.  For instance, in
\astrostep{3} we found that a chunk size of $[1000 \times 1000]$ of the LSST images leads
to the best performance.
Chunk size $[500 \times 500]$, for example, is 3$\times$ slower;
Chunk sizes $[1500 \times 1500]$ and $[2000 \times 2000]$ are slower by 22\% and 55\%, respectively.

\subsubsection{Memory Management}
Image analytics workloads are memory intensive. As we showed earlier
in \autoref{mri_datasize}, the input data for 25 subjects is more than
100GB in size. The largest intermediate dataset produced by the
analysis pipeline is over 200GB in size. Similarly, 24 exposures in
the astronomy dataset yield over 100GB in size and nearly 300GB during
processing. Additionally, data allocation can be skewed across compute
nodes. For example, the astronomy pipeline grows the data by 2.5$\times$ on
average during processing, but some workers experience data growth of
6$\times$ due to skew. As a result, image analytics pipelines can easily
experience out-of-memory failures. Big data systems can use different
approaches to trade-off query execution time and memory
consumption. We evaluate some of these trade-offs in this section. In
particular, we compare the performance of pipelined and materialized
query execution.  For materialized query execution, we compare
materializing intermediate results and processing subsets of the input
data at a time. \autoref{astro-mat} shows the results for the Myria
system on the astronomy use case. As the figure shows, when data is
small compared with the available memory, the fastest way to execute
the query is to pipeline intermediate results from one operator to the
next. This approach is 8-11\% faster in this experiment than
materialization and 15-23\% faster than executing multiple queries. As
the ratio of data-to-memory grows, pipelining may cause a query to
fail with an out-of-memory error. Intermediate result materialization
then becomes the fastest query execution method. With even more data,
the system must cut the data analysis into even smaller pieces for
execution without failure.

In contrast to Myria, Spark can spill intermediate results to disk to
avoid out-of-memory failures. Additionally, as we discussed in
\autoref{sec:optimize:parallel}, Spark can partition data into smaller
fragments than the number of available nodes, and will process only as
many fragments as there are available slots. Both techniques help to avoid
out-of-memory failures. However, this approach also causes Spark to be
slower than Myria when memory is plentiful as shown earlier in
\autoref{e2e_nodes_astro}.

\subsubsection{Data Caching}
Spark supports caching data in memory. Intermediate results replace the input data in memory as the computation proceeds and unless data is specifically cached a branch in the
computation graph may result in re-computation of intermediate data. Caching can be harmful if the
results are not needed by multiple steps as caching reduces the memory
available to query processing.  In our use cases, opportunities for data
reuse are limited. Nevertheless, we found that caching the input data
for the neuroscience use case yielded a consistent 7-8\% runtime
improvement across input data sizes.

%% file: futurework.tex
\section{Summary and Future Work}
\label{sec:futurework}

Our experiments have shown that while all of the evaluated systems
have their strengths and weaknesses, none of them serves all the needs
of large-scale image analytics.  We summarize our key lessons learned:

\textbf{User Defined Functions:} Supporting user defined functions
written in the language favored by scientists is an important
requirement of large-scale image analytics. Scientists have legacy
code that they seek to port and extend. Rewriting such computation
using the domain-specific language provided by big data systems (\eg,
SQL, AFL, {\it etc.})  is non-trivial, error-prone, and sometimes
impossible when required functions (\eg, multidimensional convolution)
are missing. Dask, Spark, and Myria were easier to use because of
their flexible support for user-defined code in Python. SciDB's {\tt
  stream()} interface is an exciting new development in this regard.

\textbf{Data Partitioning:} When porting an existing image analytics
pipeline to a big data system, the user must specify how to partition
data across the cluster before invoking legacy, user-defined
operations on the data. One challenge is that reference
implementations do not indicate how computation can be
parallelized. For example, Model building
\neurostep{3} processes the data for each voxel independently, but
this independence is not evident from the reference implementation. In
other cases, the reference implementation does not support parallelism
even though the algorithm permits it. For example, \astrostep{2}
processes entire exposures even though only subsets of pixels
are used to create each patch.  An interesting area for future work is
to enable support for legacy user code for parallel image analytics
that does not require manual specification nor tuning of the data
partitioning at each step.

\textbf{Data Formats:} Our evaluation shows the need for supporting
user-defined functions but it also shows the performance of native
operation implementations. A key challenge lies in data format
transformations between the two types of operations. Predefined
operations in big data systems work with internal formats (\eg, SciDB
arrays, Myria relations, TensorFlow tensors) but user-defined
functions use language- or domain-specific formats (\eg, NumPy arrays
or FITS files). Conversions between formats adds overhead and
complicates implementations. An interesting area of future work is to
optimize away these format conversions.

\textbf{System Tuning:} All systems needed tuning, and none of them
performed best with the default settings. System tuning, however,
requires a deeper understanding of each of the systems, which is
beyond the knowledge that should be required from users.
Self-tuning thus remains an important goal for big data systems.

%% file: relatedwork.tex
\section{Related Work}

Traditionally, image processing research has focused on effective
indexing and querying of mutli-media
content~\cite{faloutsos:94,carson:99,chaudhuri:04}. These systems
focus on utilizing image content to create indices using attributes
like color, texture, shape of image objects, and regions and then
specifying similarity measures for querying, joining, {\em etc}.

There have been many benchmarks proposed by the DBMS community over
the years such as
the TPC benchmarks~\cite{tpch}. These benchmarks, however, focus on traditional
Business Intelligence computations over structured data.
Recently, the GenBase benchmark~\cite{taft:14} took this forward to
focus on complex analytics besides data management tasks, but did not
examine image data. Pavlo \ea,~\cite{pavlo:09}
evaluated the performance of MapReduce systems against
parallel databases, but the analysis was limited to
specific text analytic queries, not image analysis.

Several big data systems (\cite{flink, impala, hadoop, rasdaman}) with
similar capabilities to the ones we evaluated are available for large
scale data analysis. We picked five representative systems that cover
today's key processing paradigms. We considered
Rasdaman~\cite{rasdaman}, which is an array database with capabilities
similar to SciDB, but were unable to make much progress as the
community version does not support UDFs.

%% file: conclusion.tex
\section{Conclusion}

We presented the first comprehensive study of large-scale image
analytics on big data systems. We surveyed the major paradigms of
large-scale data processing platforms using two real-world use cases
from domain sciences.  Our analysis shows that while we were able to
execute the use-cases on several systems, leveraging the benefits of
all systems required deep technical expertise.  As such, we argue that
current systems exhibit significant opportunities for further
improvement and future research.

\section{Acknowledgements}
This project is supported in part by the National Science Foundation through NSF
grants IIS-1247469, IIS-1110370, IIS-1546083, AST-1409547 and CNS-1563788; DARPA award
FA8750-16-2-0032; DOE award DE-SC0016260,  DE-SC0011635,
and from the DIRAC Institute, gifts from Amazon, Adobe, Google, the Intel
Science and Technology Center for Big Data, award from the Gordon and Betty
Moore Foundation and the Alfred P Sloan Foundation, and the Washington Research
Foundation Fund for Innovation in Data-Intensive Discovery.